\documentclass[twocolumn,showpacs,preprintnumbers,amsmath,amssymb,superscriptaddress,balancelastpage,prb]{revtex4}
\pdfoutput=1

\include{options}
\usepackage {epsfig,color}
\usepackage{graphicx}

\begin{document}

\title{Cooperative effect of doping and temperature on the polaronic band structure in strongly correlated electron systems with strong electron-phonon interaction}

\author{I.~A.~Makarov}
\author{S.~G.~Ovchinnikov}
	\email{sgo@iph.krasn.ru}
\affiliation{Kirensky Institute of Physics, Federal Research Center KSC SB RAS, 660036 Krasnoyarsk, Russia}

\date{\today}


\begin{abstract}

In this work we investigate doping and temperature dependences of electronic structure of system with strong electronic correlations and strong electron-phonon interaction modeling cuprates in the frameworks of the three-band p-d-Holstein model by a polaronic version of the generalized tight binding (GTB) method. Within this approach the electronic structure is formed by polaronic quasiparticles constructed as excitations between initial and final polaronic multielectron states. Doping and temperature effects are taken into account by occupation numbers of local excited polaronic states and variations in the magnitude of spin-spin correlation functions. Both effects are manifested in the reconstruction of band structure, Fermi contours, density of states and redistribution of the spectral weight over the Hubbard polaron subbands. Doping leads to transformation of Fermi contour from small hole pockets around ${\bf{k}} = \left( {{\pi  \mathord{\left/
 {\vphantom {\pi  2}} \right.
 \kern-\nulldelimiterspace} 2},{\pi  \mathord{\left/
 {\vphantom {\pi  2}} \right.
 \kern-\nulldelimiterspace} 2}} \right)$ with inhomogeneous spectral weight distribution at small hole concentration to large contour around ${\bf{k}} = \left( {\pi ,\pi } \right)$ in the overdoped compound as a result of two quantum phase transitions. In the system with phonon subsystem and EPI doping results in the top of the valence band splitting off and new polaron subbands appearance. Temperature increasing in the system with doped holes and moderate EPI leads to formation of the flatband around ${\bf{k}} = \left( {\pi ,\pi } \right)$ and transfer of the spectral weight to the splitted off top of the valence band.

\end{abstract}


\pacs{71.38.-k, 74.72.-h}

\maketitle

\section{Introduction \label{sec_introduction}}
Doping dependence of electronic structure is the main feature of HTSC cuprates. In the undoped cuprates such as Ca$_2$CuO$_2$Cl$_2$ and La$_2$CuO$_4$ ARPES measurements indicate large linewidth of the lower Hubbard band (LHB). Addition of extra holes brings to appearance of narrow coherent peak crossing Fermi level at point of $k$-space in the nodal direction. Intensity of this quasiparticle peak grows with increasing concentration of doped carriers, spectral weight is transferred from incoherent broad peak to coherent peak. Further hole concentration increasing brings to formation of arc with significant intensity as Fermi surface. Arc can be considered as part of large contour around ${\bf{k}} = \left( {\pi ,\pi } \right)$ point. Maxima of spectral functions at $k$-point along this large contour form characteristic profile, position of spectral function maximum move away from Fermi level as we go from nodal direction to antinodal direction. Position of coherent peak of the spectral function in the antinodal direction is separated by maximal pseudogap from Fermi level. Pseudogap is gradually filled for $k$-points along the contour with doping and arc transforms to large contour. In the overdoped cuprates Fermi contour is large contour around ${\bf{k}} = \left( {\pi ,\pi } \right)$.

Doping also influences on temperature dependence of ARPES spectra. Temperature dependence of spectra is strong in the undoped cuprates Ca$_2$CuO$_2$Cl$_2$, Sr$_2$CuO$_2$Cl$_2$~\cite{Kim2002} and La$_2$CuO$_4$.~\cite{ShenKM2007} Width of the ARPES spectra grows with temperature increasing, wherein position of peak is shifted deep into the band and its spectral weight is lowered. Broadening of ARPES spectra with temperature increasing was explained in the frameworks of concept of Franck-Condon broadening, polaron physics and strong EPI.~\cite{CatFilMisNag,Rosch2005} Franck-Condon broadening which is appeared in the systems with strong electron-phonon interaction (EPI) due to polaronic excitations formation. Initially Franck-Condon principle was suggested for explanation of absorption spectra of molecules.~\cite{Franck1926,Condon1926,Condon1928} The Franck-Condon principle is that probability of excitation between vibrational levels of initial and final multielectron states will be nonzero if equilibrium position of oscillating system of atoms in the final state is shifted relative to initial state. Intensity of  transition is proportional to overlap of initial and final phonon wave functions (Franck-Condon factor). And electronic transition transforms to several peaks of multiphonon excitations, intensity of every peak depends on Franck-Condon factor (overlap of phonon wave functions of the initial and final states). Finite lifetime of quasiparticles leads to transformation of quantity of multiphonon peaks merge into pne broad peak. This ideology was applied for explanation of ARPES spectra in the undoped cuprates in work Ref.~\onlinecite{ShenKM2004}.

In the overdoped cuprates temperature dependence is much weaker. Reason of temperature effect damping with doping is still unclear.

To figure out mechanism of interplay of doped holes concentration and polaron formation which leads to broadening of spectra we investigate doping and temperature dependence of the electronic structure for the three-band $p-d$-Holstein model by polaronic Generalized Tight-Binding (p-GTB) method.~\cite{Makarov2015} The system under study is compound with SEC and strong EPI the prototype of which is the undoped single-layer cuprate La$_2$CuO$_4$. In the frameworks of GTB method the electronic structure is formed by bands of the Hubbard fermion quasiparticles which are electron excitations between certain initial and final local multielectron states. Spectral weight of excitations depends on probability of transition from initial to final state and filling of these states. Structure and energy of the multielectron states have been found by exact diagonalization of the local part of the total Hamiltonian for the CuO$_6$ cluster with taking into account all Coulomb and electron-phonon interactions in the cluster. With the Hubbard $X$-operators constructed using the exact local multielectron and multiphonon eigenstates the intercluster hopping can be treated in the polaronic version of GTB. It results in the band structure of the Hubbard polarons that are formed by a hybridization of the Hubbard fermions without EPI and the local Franck-Condon resonances. In our approach concentration of doped carriers influences on electronic subsystem by occupation of the ground local multielectron two-hole state and spin subsystem by the change of magnitude of spin-spin correlations for a short-range order magnetic state.

We will study concentration dependence of pure electronic system within three-band $p-d$ model, concentration dependence of the system including phonon subsystem and moderate EPI ($\lambda=0.1$) at low ($T=10$ K) and high ($T=760$ K) temperatures.

The organization of this work is as follows. In Section~\ref{Model_and_GTB} we briefly remind a scheme of polaronic GTB method and its realization for three-band-Holstein model. Section~\ref{x_dep_without_EPI} devoted to doping effects on the electronic spectra of the pure electronic system without phonon subsystem and EPI. In Section~\ref{x_dep_with_EPI} the dependence of electronic structure on concentration of doped holes for system consisting electron and phonon subsystems and EPI is discussed. Section~\ref{T_x_dep_with_EPI} contains temperature dependence of the polaronic band structure at different concentration of doped holes. Section~\ref{Summary} is devoted to the main conclusions.

\section{Three-band $p-d$-Holstein model within polaronic GTB-method\label{Model_and_GTB}}

The GTB method was developed in the works Refs.~\cite{Ovchinnikov89,Gav2000,LDA+GTB} to calculate band structure of quasiparticle excitations in the strongly correlated materials. This method includes several steps. Crystal lattice of atoms is presented as a set of separate clusters (unit cells as a special case). The total Hamiltonian of the crystal is divided into two parts. One part is sum of Hamiltonians of separate clusters. Second part is Hamiltonian of intercluster hopping and interactions. The exact diagonalization of the Hamiltonian for a separate cluster with different number of fermions gives the multielectron eigenstates and eigenvalues of the cluster. At this stage all interactions inside the cluster are taken into account. Further Fermi-type excitations between initial and final multielectron states with $N$ and $N+1$ fermions described by the Hubbard operators are introduced. At the last stage of the GTB-method the total Hamiltonian including intercluster interactions is rewritten in the terms of Hubbard operators. Therefore the Hamiltonian of the original model is exactly converted to the multiband Hubbard model in the GTB-method.

It is widely accepted that a low-energy electronic structure of HTSC cuprates is formed by distribution of holes in the CuO$_2$ plane over copper ${d_{{x^2} - {y^2}}}$ (hereinafter $d$) and oxygen ${p_{x,y}}$ orbitals. Phonon system will be described by dispersionless local vibrations of breathing mode, light O atoms vibrate relative to the fixed heavy Cu atoms. Displacement of O atoms from its equilibrium positions changes value of the crystal field for the hole on Cu atom and the hopping integral between copper and oxygen atomic orbitals. In general the EPI includes a diagonal part which is written as renormalization of on-site energy on Cu atom and an off-diagonal EPI which is defined by renormalization of the $p-d$ hopping integral. Evolution of the structure of the local polaronic states and band structure of polaronic quasiparticles with varying diagonal and off-diagonal EPI was discussed in Refs.~\onlinecite{Makarov2015,Makarov2016}. Ratio between diagonal and off-diagonal EPI influences only on character of transition, smooth or abrupt, between states with weak and strong localization (a measure of localization is the average number of holes on $d$-orbital) with varying EPI parameters. In this work we consider only the diagonal EPI. Thus we will use the three-band $p-d$-Holstein model:
\begin{eqnarray}
\label{pd_Hamiltonian}
H & = & {H_{el}} + {H_{ph}} + {H_{e - ph}} \nonumber \\
{H_{el}} & = & \sum\limits_{\bf{f}\sigma } {{\varepsilon _d}d_{\bf{f}\sigma }^\dag {d_{\bf{f}\sigma }}}  + \sum\limits_{\alpha \bf{h}\sigma } {{\varepsilon _p}p_{\alpha\bf{h}\sigma }^\dag {p_{\alpha \bf{h}\sigma }}}  + \nonumber \\
& & + \sum\limits_{\bf{fh}\alpha \sigma } {{{\left( { - 1} \right)}^{{R_{\bf{h}}}}}{t_{pd}}\left( {d_{\bf{f}\sigma }^ \dag {p_{\alpha\bf{h}\sigma }} + h.c.} \right)}  + \nonumber \\
& & + \sum\limits_{\alpha \alpha '\bf{h} \ne \bf{h'}\sigma } {{{\left( { - 1} \right)}^{{M_{\bf{hh'}}}}}{t_{pp}}\left( {p_{\alpha\bf{h}\sigma }^\dag {p_{{\alpha '}\bf{h'}\sigma }} + h.c.} \right)}  + \nonumber \\
& & + \sum\limits_{\bf{f}} {{U_d}d_{\bf{f} \uparrow }^\dag {d_{\bf{f} \uparrow }}d_{\bf{f} \downarrow }^\dag {d_{\bf{f} \downarrow }}}  + \nonumber \\
& & + \sum\limits_{\bf{h}} {{U_p}p_{\alpha\bf{h} \uparrow }^\dag {p_{\alpha\bf{h} \uparrow }}p_{\alpha\bf{h} \downarrow }^\dag {p_{\alpha\bf{h} \downarrow }}}  + \nonumber \\
& & + \sum\limits_{\alpha\bf{fh}\sigma \sigma '} {{V_{pd}}d_{\bf{f}\sigma }^\dag {d_{\bf{f}\sigma }}p_{\alpha\bf{h}\sigma '}^\dag {p_{\alpha\bf{h}\sigma '}}} \nonumber \\
{H_{ph}} & = &\frac{M}{2}\sum\limits_{\bf{h}} {\left( {\dot u_{\bf{h}}^2 + \omega _b^2\dot u_{\bf{h}}^2} \right)} \nonumber \\
{H_{e - ph}} & = & \sum\limits_{\bf{f}\sigma } {{\left( {\sum\limits_{\bf{h}} {{{\left( { - 1} \right)}^{{S_{\bf{h}}}}}{g_d}{u_{\bf{h}}}} } \right)} d_{\bf{f}\sigma }^\dag {d_{\bf{f}\sigma }}}
\end{eqnarray}
Here ${d_{\bf{f}\sigma }}$ and ${p_{\alpha\bf{h}\sigma }}$ are the operators of hole annihilation with spin $\sigma $ on $d$-orbital of the copper atom ${\bf{f}}$ and ${p_x}$(${p_y}$) -orbital of the oxygen atom ${\bf{h}}$, respectively. ${\bf{h}}$ runs over two of the four positions of planar oxygen atoms neighboring to Cu atom in octahedral unit cell centered on site ${\bf{f}}$ at each $\alpha$, ${\bf{h}} = \left\{ {\left( {{f_x} - {a \mathord{\left/
 {\vphantom {a 2}} \right.
 \kern-\nulldelimiterspace} 2},{f_y}} \right),\left( {{f_x} + {a \mathord{\left/
 {\vphantom {a 2}} \right.
 \kern-\nulldelimiterspace} 2},{f_y}} \right)} \right\}$ if $\alpha=x$ and ${\bf{h}} = \left\{ {\left( {{f_x},{f_y} - {b \mathord{\left/
 {\vphantom {b 2}} \right.
 \kern-\nulldelimiterspace} 2}} \right),\left( {{f_x},{f_y} + {b \mathord{\left/
 {\vphantom {b 2}} \right.
 \kern-\nulldelimiterspace} 2}} \right)} \right\}$ if $\alpha=y$, $a$ and $b$ are the lattice parameters. ${\varepsilon _d}$ is the on-site energy of hole on Cu ion and ${\varepsilon _p}$ is the same on O ion; $t_{pd}$ is the amplitude of nearest-neighbor hopping between $d$-orbitals of Cu ion ${\bf{f}}$ and ${p_{x,y}}$-orbitals of O ion ${\bf{h}}$ in CuO$_2$ plane and $t_{pp}$ is the amplitude of nearest-neighbor hopping between ${p_{x,y}}$-orbitals of the oxygen atoms ${\bf{h}}$ and ${\bf{h'}}$. The phase parameters ${R_{\bf{h}}}$ and ${M_{\bf{hh'}} }$ are determined by phases of overlapping wave functions. ${U_d}$ is the Coulomb interaction of two holes on the same copper atom and ${U_p}$ is the same for oxygen atom, $V_{pd}$ is the intersite Coulomb interaction when one hole is on the copper orbital of ${\bf{f}}$ atom and other hole is on the oxygen orbital of ${\bf{h}}$ atom. ${u_{\bf{h}}} = \sqrt {\frac{\hbar }{{2M{\omega _b}}}} \left( {e_{\alpha\bf{h}}^ \dag  + {e_{\alpha\bf{h}}}} \right)$ is the operator of oxygen atom ${\bf{h}}$ displacement, $M$ is the mass of oxygen atom. $e_{\alpha\bf{h}}^\dag $ is the operator of creation of local phonon with frequency ${\omega _b}$, $\alpha$ denotes direction of atom ${\bf{h}}$ displacement. For oxygen atom ${\bf{h}} = \left\{ {\left( {{f_x} - {a \mathord{\left/
 {\vphantom {a 2}} \right.
 \kern-\nulldelimiterspace} 2},{f_y}} \right),\left( {{f_x} + {a \mathord{\left/
 {\vphantom {a 2}} \right.
 \kern-\nulldelimiterspace} 2},{f_y}} \right)} \right\}$ (${\bf{h}} = \left\{ {\left( {{f_x},{f_y} - {b \mathord{\left/
 {\vphantom {b 2}} \right.
 \kern-\nulldelimiterspace} 2}} \right),\left( {{f_x},{f_y} + {b \mathord{\left/
 {\vphantom {b 2}} \right.
 \kern-\nulldelimiterspace} 2}} \right)} \right\}$) displacement is along $\alpha=x$ ($y$) axis. ${g_d}$ is the parameter of the diagonal EPI between hole, located on copper atom and phonon of breathing mode. The phase parameter ${S_{\bf{h}} } = 0$ for ${\bf{h}} = \left( {{f_x} + {a \mathord{\left/
 {\vphantom {a 2}} \right.
 \kern-\nulldelimiterspace} 2},{f_y}} \right),\left( {{f_x},{f_y} + {b \mathord{\left/
 {\vphantom {b 2}} \right.
 \kern-\nulldelimiterspace} 2}} \right)$ and ${S_{\bf{h}} } = 1$ for ${\bf{h}} = \left( {{f_x} - {a \mathord{\left/
 {\vphantom {a 2}} \right.
 \kern-\nulldelimiterspace} 2},{f_y}} \right),\left( {{f_x},{f_y} - {b \mathord{\left/
 {\vphantom {b 2}} \right.
 \kern-\nulldelimiterspace} 2}} \right)$, it is consistent with modulation of the on-site energy for the breathing mode. We introduce dimensionless EPI parameters ${\lambda _{d}} = {{{{\left( {{g_{d}}\xi } \right)}^2}} \mathord{\left/
 {\vphantom {{{{\left( {{g_{d}}\xi } \right)}^2}} {W\hbar {\omega _b}}}} \right.
 \kern-\nulldelimiterspace} {W\hbar {\omega _b}}}$, where $\xi  = \sqrt {\frac{\hbar }{{2M{\omega _b}}}} $, $W$ is the bandwidth of the free electron in tight-binding method without EPI, we accept here $W = 1$ eV and the breathing mode phonon energy $\hbar {\omega _b} = 0.04$ eV.

Below we will calculate the quasiparticle band structure with the following parameters of Hamiltonian (in eV):
\begin{eqnarray}
{\varepsilon _d} = 0, {\varepsilon _p} = 1.5, {t_{pd}} = 1.36, {t_{pp}} = 0.86 \nonumber \\
{U_d} = 9, {U_p} = 4, {V_{pd}} = 1.5
\label{realistic_parameters}
\end{eqnarray}
calculated for La$_2$CuO$_4$ from \textit{ab-initio} approach.~\cite{LDA+GTB}

Since each planar oxygen atom belongs to the two CuO$_6$ clusters at once we should make orthogonalization procedure. We proceed from the hole atomic oxygen orbitals $\left| {{p_{x{\bf{h}}}}} \right\rangle $ and $\left| {{p_{y{\bf{h}}}}} \right\rangle $ to the molecular oxygen orbitals $\left| {{b_{\bf{f}}}} \right\rangle $ and $\left| {{a_{\bf{f}}}} \right\rangle $ by transformation in the $k$-space:~\cite{Shastry89}
\begin{eqnarray}
\label{Shastry}
{b_{\bf{k}}}&=& \frac{i}{{{\mu _{\bf{k}}}}}\left( {{s_{{\bf{k}}x}}{p_{x{\bf{k}}}} - {s_{{\bf{k}}y}}{p_{y{\bf{k}}}}} \right)\nonumber \\
{a_{\bf{k}}}&=& - \frac{i}{{{\mu _{\bf{k}}}}}\left( {{s_{{\bf{k}}y}}{p_{x{\bf{k}}}} + {s_{{\bf{k}}x}}{p_{y{\bf{k}}}}} \right)
\end{eqnarray}
where ${s_{{\bf{k}}x}} = \sin \left( {{{{k_x}a} \mathord{\left/
 {\vphantom {{{k_x}a} 2}} \right.
 \kern-\nulldelimiterspace} 2}} \right)$, ${s_{{\bf{k}}y}} = \sin \left( {{{{k_y}b} \mathord{\left/
 {\vphantom {{{k_y}b} 2}} \right.
 \kern-\nulldelimiterspace} 2}} \right)$ and ${\mu _{\bf{k}}} = \sqrt {s_{{\bf{k}}x}^2 + s_{{\bf{k}}y}^2} $. States ${b_{\bf{f}}}$(${a_{\bf{f}}}$) in the neighbor CuO$_6$ clusters are orthogonal to each other. Similar transformation of the phonon states is undertaken. States of atomic phonons $e_{\bf{h}}^\dag \left| 0 \right\rangle $ are replaced by "molecular" oxygen phonon wave functions $A_{\bf{f}}^\dag \left| 0 \right\rangle $ and $B_{\bf{f}}^\dag \left| 0 \right\rangle $, where operators ${A_{\bf{k}}}$ and ${B_{\bf{k}}}$ in the momentum space are defined by transformation:

\begin{eqnarray}
\label{PhOper}
{A_{\bf{k}}} =  - \frac{i}{{{\mu _{\bf{k}}}}}\left( {{s_{{\bf{k}}x}}{e_{x{\bf{k}}}} + {s_{{\bf{k}}y}}{e_{y{\bf{k}}}}} \right)\nonumber \\
{B_{\bf{k}}} =  - \frac{i}{{{\mu _{\bf{k}}}}}\left( {{s_{{\bf{k}}y}}{e_{x{\bf{k}}}} - {s_{{\bf{k}}x}}{e_{y{\bf{k}}}}} \right)
\end{eqnarray}
Phonon wave functions $B_{\bf{f}}^\dag \left| 0 \right\rangle $ have negligible contribution to the low-energy local cluster eigenstates and consequently to the electron spectral function of the top of the valence band and the bottom of the conduction band. Therefore we neglect them in the further consideration taking into account only $A$-mode phonons.

After orthogonalization procedure the Hamiltonian of three-band $p-d$-Holstein model can be written as
\begin{eqnarray}
\label{Full_Ham}
& H & = \sum\limits_{{\bf{f}}\sigma } {{\varepsilon _d}d_{{\bf{f}}\sigma }^\dag {d_{{\bf{f}}\sigma }}}  + \sum\limits_{{\bf{f}}\sigma } {{\varepsilon _p}b_{{\bf{f}}\sigma }^\dag {b_{{\bf{f}}\sigma }}}  - \nonumber \\
& & - 2{t_{pd}}\sum\limits_{{\bf{fg}}\sigma } {{\mu _{\bf{fg}}}\left( {d_{{\bf{f}}\sigma }^\dag {b_{{\bf{g}}\sigma }} + h.c.} \right)}  - \nonumber \\
& & - 2{t_{pp}}\sum\limits_{{\bf{fg}}\sigma } {{\nu _{\bf{fg}}}\left( {b_{{\bf{f}}\sigma }^\dag {b_{{\bf{g}}\sigma }} + h.c.} \right)}  +\nonumber \\
& & + \sum\limits_{\bf{f}} {{U_d}d_{{\bf{f}} \uparrow }^\dag {d_{{\bf{f}} \uparrow }}d_{{\bf{f}} \downarrow }^\dag {d_{{\bf{f}} \downarrow }}}  + \nonumber \\
& & + \sum\limits_{\bf{fghl}} {{U_p}{\Psi _{\bf{fghl}}}b_{{\bf{f}} \uparrow }^\dag {b_{{\bf{g}} \uparrow }}b_{{\bf{h}} \downarrow }^\dag {b_{{\bf{l}} \downarrow }}}  + \nonumber \\
& & + \sum\limits_{{\bf{fgh}}\sigma \sigma '} {{V_{pd}}{\Phi _{\bf{fgh}}}d_{{\bf{f}}\sigma }^\dag {d_{{\bf{f}}\sigma }}b_{{\bf{g}}\sigma '}^\dag {b_{{\bf{h}}\sigma '}}}  + \nonumber \\
& & + \sum\limits_{\bf{f}} {\hbar {\omega _b}A_{\bf{f}}^\dag {A_{\bf{f}}}}  + \nonumber \\
& & + \sum\limits_{\bf{fg}} {2{g_d}{\xi _d}{\mu _{\bf{fg}}}\sum\limits_\sigma  {\left( {A_{\bf{f}}^\dag  + {A_{\bf{f}}}} \right)d_{{\bf{g}}\sigma }^\dag {d_{{\bf{g}}\sigma }}} }
\end{eqnarray}
where structural factors ${\mu _{\bf{fg}}}$, ${\nu _{\bf{fg}}}$, ${\rho _{\bf{fgh}}}$ are defined as
\begin{eqnarray}
\label{str_factors}
{\mu _{\bf{fg}}} & = & {1 \mathord{\left/
 {\vphantom {1 N}} \right.
 \kern-\nulldelimiterspace} N}\sum\limits_{\bf{k}} {{\mu _{\bf{k}}}{e^{ - i{\bf{k}}\left( {{\bf{f}} - {\bf{g}}} \right)}}} \nonumber \\
{\nu _{\bf{fg}}} & = & {1 \mathord{\left/
 {\vphantom {1 N}} \right.
 \kern-\nulldelimiterspace} N}\sum\limits_{\bf{k}} {{{\left( {{{2\sin \left( {{{{k_x}} \mathord{\left/
 {\vphantom {{{k_x}} 2}} \right.
 \kern-\nulldelimiterspace} 2}} \right)\sin \left( {{{{k_y}} \mathord{\left/
 {\vphantom {{{k_y}} 2}} \right.
 \kern-\nulldelimiterspace} 2}} \right)} \mathord{\left/
 {\vphantom {{2\sin \left( {{{{k_x}} \mathord{\left/
 {\vphantom {{{k_x}} 2}} \right.
 \kern-\nulldelimiterspace} 2}} \right)\sin \left( {{{{k_y}} \mathord{\left/
 {\vphantom {{{k_y}} 2}} \right.
 \kern-\nulldelimiterspace} 2}} \right)} {{\mu _{\bf{k}}}}}} \right.
 \kern-\nulldelimiterspace} {{\mu _{\bf{k}}}}}} \right)}^2}{e^{ - i{\bf{k}}\left( {{\bf{f}} - {\bf{g}}} \right)}}} \nonumber \\
\rho _{{\bf{fgh}}}^A & = &\rho _{{\bf{f}} - {\bf{g}},{\bf{g}} - {\bf{h}}}^A{{ = 1} \mathord{\left/
 {\vphantom {{ = 1} {{N^2}}}} \right.
 \kern-\nulldelimiterspace} {{N^2}}}\sum\limits_{{\bf{kq}}} {{1 \mathord{\left/
 {\vphantom {1 {{\mu _{\bf{k}}}{\mu _{\bf{q}}}}}} \right.
 \kern-\nulldelimiterspace} {{\mu _{\bf{k}}}{\mu _{\bf{q}}}}}}  \times \nonumber \\
& \times & \left[ {\sin \left( {{{{k_x}} \mathord{\left/
 {\vphantom {{{k_x}} 2}} \right.
 \kern-\nulldelimiterspace} 2}} \right)\sin \left( {{{{q_x}} \mathord{\left/
 {\vphantom {{{q_x}} 2}} \right.
 \kern-\nulldelimiterspace} 2}} \right)\cos \left( {{{\left( {{k_x} + {q_x}} \right)} \mathord{\left/
 {\vphantom {{\left( {{k_x} + {q_x}} \right)} 2}} \right.
 \kern-\nulldelimiterspace} 2}} \right) + } \right.\nonumber \\
& + & \left. {\sin \left( {{{{k_y}} \mathord{\left/
 {\vphantom {{{k_y}} 2}} \right.
 \kern-\nulldelimiterspace} 2}} \right)\sin \left( {{{{q_y}} \mathord{\left/
 {\vphantom {{{q_y}} 2}} \right.
 \kern-\nulldelimiterspace} 2}} \right)\cos \left( {{{\left( {{k_y} + {q_y}} \right)} \mathord{\left/
 {\vphantom {{\left( {{k_y} + {q_y}} \right)} 2}} \right.
 \kern-\nulldelimiterspace} 2}} \right)} \right] \times \nonumber \\
& \times & {e^{ - i{\bf{k}}\left( {{\bf{f}} - {\bf{g}}} \right)}}{e^{ - i{\bf{q}}\left( {{\bf{g}} - {\bf{h}}} \right)}}
\end{eqnarray}
Values of coefficients ${\Psi _{\bf{fghl}}}$ and ${\Phi _{\bf{fgh}}}$ strongly decrease with increasing distance between sites ${\bf{f,g,h}}$ ~\cite{Feiner1996}, therefore we restrict our consideration by only intracluster Coulomb interactions, ${\Psi _{0000}} = 0.2109$, ${\Phi _{000}} = 0.918$. Now we can divide Hamiltonian $H$ on the intracluster part and intercluster interactions:
\begin{equation}
\label{Hc_Hcc_tot}
H = {H_c} + {H_{cc}}, {H_c} = \sum\limits_{\bf{f}} {{H_{\bf{f}}}}, {H_{cc}} = \sum\limits_{\bf{fg}} {{H_{\bf{fg}}}}
\end{equation}

\begin{figure}
\center
\includegraphics[width=1.0\linewidth]{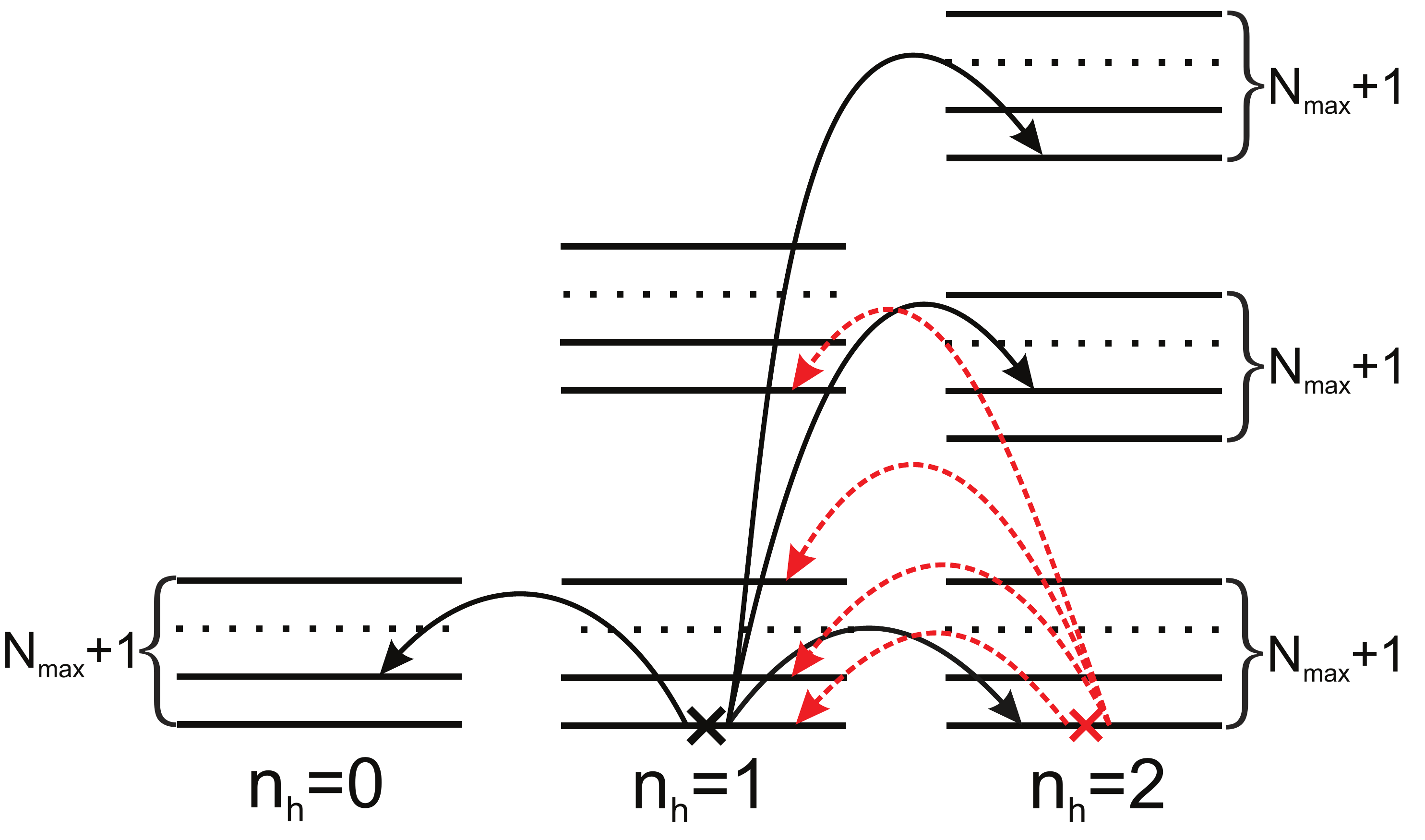}
\caption{\label{fig:levels} Schematic picture of the multielectron vibronic eigenstates of CuO$_6$ cluster (black horizontal lines) with hole numbers ${n_h} = 0,1,2$ and Fermi-type excitations between them. Black solid arrows denotes excitations with nonzero spectral weight at $T = 0$ K due to occupation of single-hole ground state (black cross) for undoped La$_2$CuO$_4$. At nonzero doping ground two-hole state are occupied (red crosses) and excitations involving this state acquire spectral weight (red dashed arrows).}
\end{figure}

Eigenstates of the CuO$_6$ cluster with hole numbers ${n_h} = 0,1,2$ obtained by exact diagonalization of Hamiltonian ${H_c}$ include hole and phonon basis wave functions. In the hole vacuum sector of the Hilbert space with ${n_h} = 0$ eigenstates are just the harmonic oscillator states:
\begin{equation}
|0,\nu \rangle  = |0\rangle |\nu \rangle, \nu  = 0,1,...,{N_{\max }}
\label{es0}
\end{equation}
Here $|0\rangle $ denotes electronic configuration $\left| {3{d^{10}}2{p^6}} \right\rangle $, and $|\nu \rangle $ - phonon state with phonon number ${n_{ph}} = \nu $. Phonon states $\left| \nu  \right\rangle $ are $\nu $-times action of phonon creation operator ${A^\dag }$ on vacuum state   of harmonic oscillator:
\begin{equation}
\left| \nu  \right\rangle  = \frac{1}{{\sqrt {\nu !} }}{\left( {{A^\dag }} \right)^\nu }\left| {0,0,...,0} \right\rangle
\label{harm_osc}
\end{equation}
 ${N_{\max }}$ is the cutoff for number of phonons, it is calculated for each certain set of parameters from the following condition: addition of phonon number above ${N_{\max }}$, $N > {N_{\max }}$, does not change the electron spectral function. The value ${N_{\max }}$ mainly depends on the EPI coupling parameter, for example at ${\lambda _d} = 0.3$ ${N_{\max }} = 30$ for the vacuum and single-hole sectors of the Hilbert space and ${N_{\max }} = 50$ for the two-hole sector.~\cite{Makarov2015} Thus there are ${N_{\max }} + 1$ states of thermal phonons in the zero-hole sector, the oxygen atoms oscillate about their equilibrium position in the potential of crystal lattice.

The single-hole (${n_h} = 1$) cluster states can be written as
\begin{equation}
\left| {1\sigma ,i} \right\rangle  = \sum\limits_{\nu  = 0}^{N_{max}} {\left( {c_{i\nu }^d\left| {{d_\sigma }} \right\rangle \left| \nu  \right\rangle  + c_{i\nu}^b\left| {{b_\sigma }} \right\rangle \left| \nu  \right\rangle } \right)}
\label{es1}
\end{equation}
Here $\left| {{d_\sigma }} \right\rangle  = d_{{x^2} - {y^2}\sigma }^\dag \left| 0 \right\rangle $,  $\left| {{b_\sigma }} \right\rangle  = b_\sigma ^\dag \left| 0 \right\rangle $, index $i$ numerates the ground and excited single-hole eigenstates. Corresponding electronic configurations of stoichiometric La$_2$CuO$_4$ are $\left| {3{d^9}2{p^6}} \right\rangle$ and $\left| {3{d^{10}}2{p^5}} \right\rangle $.

Electron part of the two-hole (${n_h} = 2$) basis consists of the Zhang-Rice singlet state $\left| {{\rm{ZR}}} \right\rangle  = \left| {\frac{1}{{\sqrt 2 }}\left( {{d_ \downarrow }{b_ \uparrow } - {d_ \uparrow }{b_ \downarrow }} \right)} \right\rangle $, the two holes on copper ion state $\left| {{d_ \downarrow }{d_ \uparrow }} \right\rangle $ and the two holes on oxygen ion state $\left| {{b_ \downarrow }{b_ \uparrow }} \right\rangle $.The electronic configurations $\left| {3{d^9}2{p^5}} \right\rangle $, $\left| {3{d^8}2{p^6}} \right\rangle $, $\left| {3{d^{10}}2{p^4}} \right\rangle $ correspond to these states. Two-hole eigenstates of CuO$_6$ cluster with phonons are
\begin{eqnarray}
\left| {2,j} \right\rangle & = & \sum\limits_{\nu  = 0}^{N_{max}} \left(  c_{j\nu }^{ZR}\left| {{\rm{ZR}}} \right\rangle \left| \nu  \right\rangle + \right. \nonumber \\
 &+ & \left. c_{j\nu }^{dd}\left| {{d_{\downarrow }}{d_{\uparrow }}} \right\rangle \left| \nu  \right\rangle  + c_{j\nu }^{bb}\left| {{b_ \downarrow }{b_ \uparrow }} \right\rangle \left| \nu  \right\rangle \right)
\label{es2}
\end{eqnarray}
It is seen from Eq.~(\ref{es1}) and Eq.~(\ref{es2}) that each hole state of the pure electronic system transforms into ${N_{\max }} + 1$ vibronic states with taking into account phonon subsystem. A scheme of multielectron and multiphonon CuO$_6$ cluster levels (\ref{es0}), (\ref{es1}) and (\ref{es2}) is depicted in Fig.~\ref{fig:levels}. Without EPI all states are Cartesian production of hole and phonon wave functions and their energy levels are equidistant. With taking into account the EPI the equilibrium positions of oxygen atoms shift to new positions. Displacements of oxygen environment of copper atom are expressed by cloud of virtual phonons. Cluster eigenstates become polaron states i.e. they are superpositions of products of hole and phonon wave functions. Similar description of the local polaronic states has been discussed previously.~\cite{Piekarz1999} The number of the virtual phonon states in the cloud and its dependence on the EPI coupling has been discussed in detail for zero temperature in our paper Ref.~\onlinecite{Makarov2015}.

\begin{figure}
\center
\includegraphics[width=1.0\linewidth]{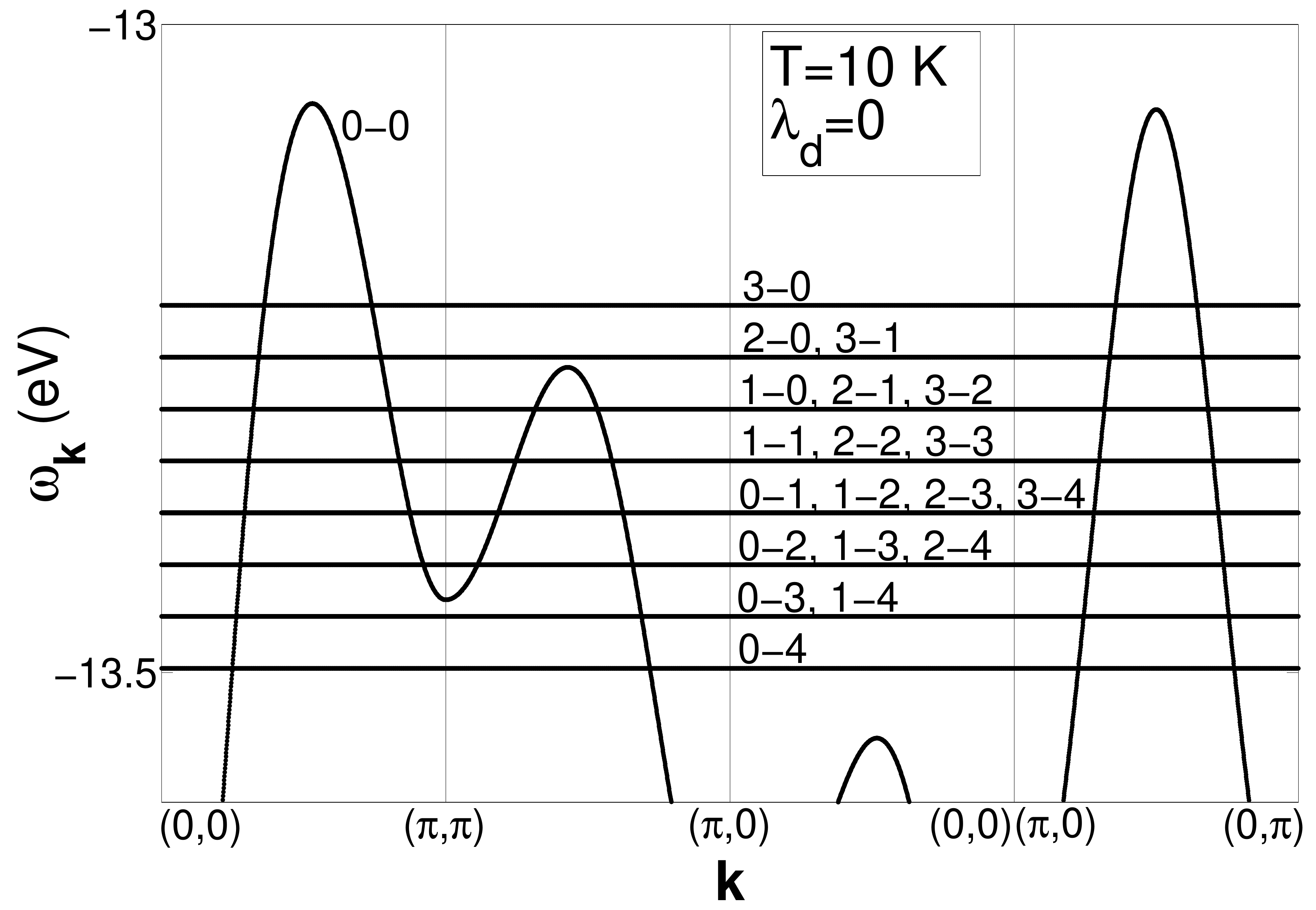}
\caption{\label{fig:FC_resonances} Dispersion of the upper part of the electron LHB without EPI at low temperature. In general the quasiparticle bands are characterized by spectral weight at each $k$-point but in this figure spectral weight difference is not shown. The quasparticle band $0-0$ (phononless excitation between ground single-hole and ground two-hole states) is dispersive due to the intercluster hopping. The Fermi-type multiphonon excitations (Franck-Condon resonances) are shown by a set of horizontal lines. They are dispersionless, multiply degenerate and has zero spectral weight at zero temperature and zero EPI constant. First (second) index in the notation of the quasiparticle denotes simultaneously index of single-hole (two-hole) cluster eigenstate.}
\end{figure}

The Fermi-type polaronic excitations (arrows in Fig.~\ref{fig:levels}) from cluster eigenstates $\left| {1,i} \right\rangle $ to $\left| {2,j} \right\rangle $ result in creation of hole, and from $\left| {1,i} \right\rangle $ to $\left| {0,\nu } \right\rangle $ results in creation of electron. Both change the number of fermions by unity, number of thermal phonons and polarization of oxygen environment of copper atom. These local polaron multiphonon excitations describe the Franck-Condon processes. Each excitation can be described by the Hubbard operator $X_{\bf{f}}^{pq} = \left| p \right\rangle \left\langle q \right|$, where $\left| q \right\rangle $ is the initial cluster eigenstate and $\left| p \right\rangle $ is the final cluster eigenstate. The Fermi-type operators of annihilation of hole on copper and oxygen orbital can be expressed in terms of the Fermi-type Hubbard operators $X_{\bf{f}}^{pq}$:
\begin{eqnarray}
{d_{{\bf{f}}\sigma }} = \sum\limits_{pq} {{\gamma _{d\sigma }}\left( {pq} \right)} X_{\bf{f}}^{pq} \nonumber \\
{b_{{\bf{f}}\sigma }} = \sum\limits_{pq} {{\gamma _{b\sigma }}\left( {pq} \right)} X_{\bf{f}}^{pq}
\label{Xop}
\end{eqnarray}
The phonon annihilation operator is expressed through the Bose-type Hubbard operators $Z_{\bf{f}}^{pp'}$:
\begin{equation}
{A_{\bf{f}}} = \sum\limits_{pp'} {{\gamma _A}\left( {pp'} \right)} Z_{\bf{f}}^{pp'}
\label{Zop}
\end{equation}
where states $\left| p \right\rangle $ and $\left| {p'} \right\rangle $ have the same number of holes and belong to the same sector of Hilbert space $\left| {{n_h}} \right\rangle $. Electronic excitations in the CuO$_6$ cluster are now described not by holes on $d$ or $b$ orbitals but polaronic quasiparticles with spectral weight defined by occupation of initial and final states of corresponding transition and their overlap. Since we will study the electronic states close to the top of the valence band that in our model is the lower Hubbard band, the Franck-Condon processes will be denoted by "index of single-hole eigenstate - index of two-hole eigenstate", $i - j$. Franck-Condon resonances are multiply degenerate, for instance single-phonon excitations $0-1$, $1-2$, $2-3$ etc have the same energy~(Fig.~\ref{fig:FC_resonances}).

Hamiltonians ${H_c}$ and ${H_{cc}}$ in Eq.(\ref{Hc_Hcc_tot}) are rewritten in the terms of Hubbard operators:
\begin{eqnarray}
& {H_c} & = \sum\limits_{\bf{f}} {\left[ {\sum\limits_l {{\varepsilon _{0l}}Z_{\bf{f}}^{0l,0l}}  + \sum\limits_i {{\varepsilon _{1i}}Z_{\bf{f}}^{1i,1i}}  + \sum\limits_j {{\varepsilon _{2j}}Z_{\bf{f}}^{2j,2j}} } \right]} \nonumber \\
& {H_{cc}} & = \sum\limits_{{\bf{f}} \ne {\bf{g}}} {\left[ {\sum\limits_{mn} {2{t_{pd}}{\mu _{{\bf{fg}}}}\gamma _{{d_x}}^ * \left( m \right){\gamma _b}\left( n \right)\mathop {X_{\bf{f}}^m}\limits^\dag  X_{\bf{g}}^n} } \right.}  - \nonumber \\
& &\left. { - \sum\limits_{mn} {2{t_{pp}}{\nu _{{\bf{fg}}}}\gamma _b^ * \left( m \right){\gamma _b}\left( n \right)\mathop {X_{\bf{f}}^m}\limits^\dag  X_{\bf{g}}^n} } \right]
\label{Hc_Hcc_Xop}
\end{eqnarray}
Here ${\varepsilon _{0l}}$, ${\varepsilon _{1i}}$, ${\varepsilon _{2j}}$ are the energies of cluster eigenstates with ${n_h} = 0,1,2$. The intercluster interactions result from the $p-d$ and $p-p$ hoppings of the Hubbard polarons between clusters, we consider hopping up to sixth neighbors. To obtain the band dispersion and spectral function of Hubbard polarons we use the equation of motion for the Green function ${D^{mn}}\left( {{\bf{f}},{\bf{g}}} \right) = \left\langle {\left\langle {{X_{\bf{f}}^m}}
 \mathrel{\left | {\vphantom {{X_{\bf{f}}^m} {X_{\bf{g}}^n}}}
 \right. \kern-\nulldelimiterspace}
\mathop {X_{\bf{g}}^n}\limits^\dag \right\rangle } \right\rangle $, where $m$,$n$ are the quasiparticle band indexes, this index is uniquely defined by initial and final states of excitation $m \equiv \left( {p,q} \right)$. The electron Green function ${G_{\lambda \lambda '}}\left( {{\bf{f}},{\bf{g}}} \right) = \left\langle {\left\langle {{{a_{\lambda {\bf{f}}}}}}
 \mathrel{\left | {\vphantom {{{a_{\lambda {\bf{}}f}}} {a_{\lambda '{\bf{g}}}^\dag }}}
 \right. \kern-\nulldelimiterspace}
 {{a_{\lambda '{\bf{g}}}^\dag }} \right\rangle } \right\rangle $ is connected with the quasiparticle Green function ${D^{mn}}\left( {{\bf{f}},{\bf{g}}} \right)$ by the following relation:
\begin{equation}
{G_{\lambda \lambda '}}\left( {{\bf{f}},{\bf{g}}} \right) = \sum\limits_{mn} {\gamma _{\lambda '}^ * \left( n \right){\gamma _\lambda }\left( m \right){D^{mn}}\left( {{\bf{f}},{\bf{g}}} \right)}
\label{GreenFunc_G_D}
\end{equation}
There are many quasiparticle excitations in our system therefore it is convenient to introduce matrix Green function $\hat D\left( {{\bf{f}},{\bf{g}}} \right)$ with matrix elements ${D^{mn}}\left( {{\bf{f}},{\bf{g}}} \right)$, indexes of row $m$ and column $n$ run over all quasiparticle bands. The set of equations of motion is decoupled in the generalized Hartri-Fock approximation by method of irreducible Green functions~\cite{Plakida1989,Yushankhay1991,Valkov2005,KorshOvch2007} taking into account the interatomic spin-spin correlation functions. The Dyson equation for the matrix Green function $\hat D\left( {{\bf{f}},{\bf{g}}} \right)$ in the momentum space has the form
\begin{equation}
\hat D\left( {{\bf{k}};\omega } \right) = {\left[ {\hat D_0^{ - 1}\left( \omega  \right) - \hat F\hat {\tilde t}\left( {{\bf{k}};\omega } \right) + \hat \Sigma \left( {{\bf{k}};\omega } \right)} \right]^{ - 1}}\hat F
\label{Dyson_eq}
\end{equation}
In this equation ${\hat D_0}$ is the exact local Green function, its matrix elements $D_0^{mn} = {{{\delta _{mn}}F\left( m \right)} \mathord{\left/
 {\vphantom {{{\delta _{mn}}F\left( m \right)} {\left( {\omega  - \Omega \left( m \right)} \right)}}} \right.
 \kern-\nulldelimiterspace} {\left( {\omega  - \Omega \left( m \right)} \right)}}$, $\Omega \left( m \right) = \Omega \left( {pq} \right) = {E_p} - {E_q}$ is the energy of the quasiparticle excitation between states $q$ and $p$, matrix elements $F^{mn} = F\left( m \right){\delta _{mn}}$, $F\left( m \right) = F\left( {pq} \right) = \left\langle {{X^{pp}}} \right\rangle  + \left\langle {{X^{qq}}} \right\rangle $ is the filling factor of the quasiparticle. The matrix of intercluster hopping $\hat {\tilde t}\left( {{\bf{k}};\omega } \right)$ is determined by $p-d$ and $p-p$ hoppings with the matrix elements $\tilde t_{\bf{k}}^{mn} = \sum\limits_{\lambda \lambda '} {\gamma _\lambda ^ * \left( m \right)} {\gamma _{\lambda '}}\left( n \right)\tilde t_{\bf{k}}^{\lambda \lambda '}$. $\hat \Sigma \left( {{\bf{k}};\omega } \right)$ is the self-energy operator which contains spin-spin correlation functions. The filling factor results in the temperature dependence of the quasiparticle band dispersion and the spectral weight.
\begin{figure*}
\center
\includegraphics[width=0.45\linewidth]{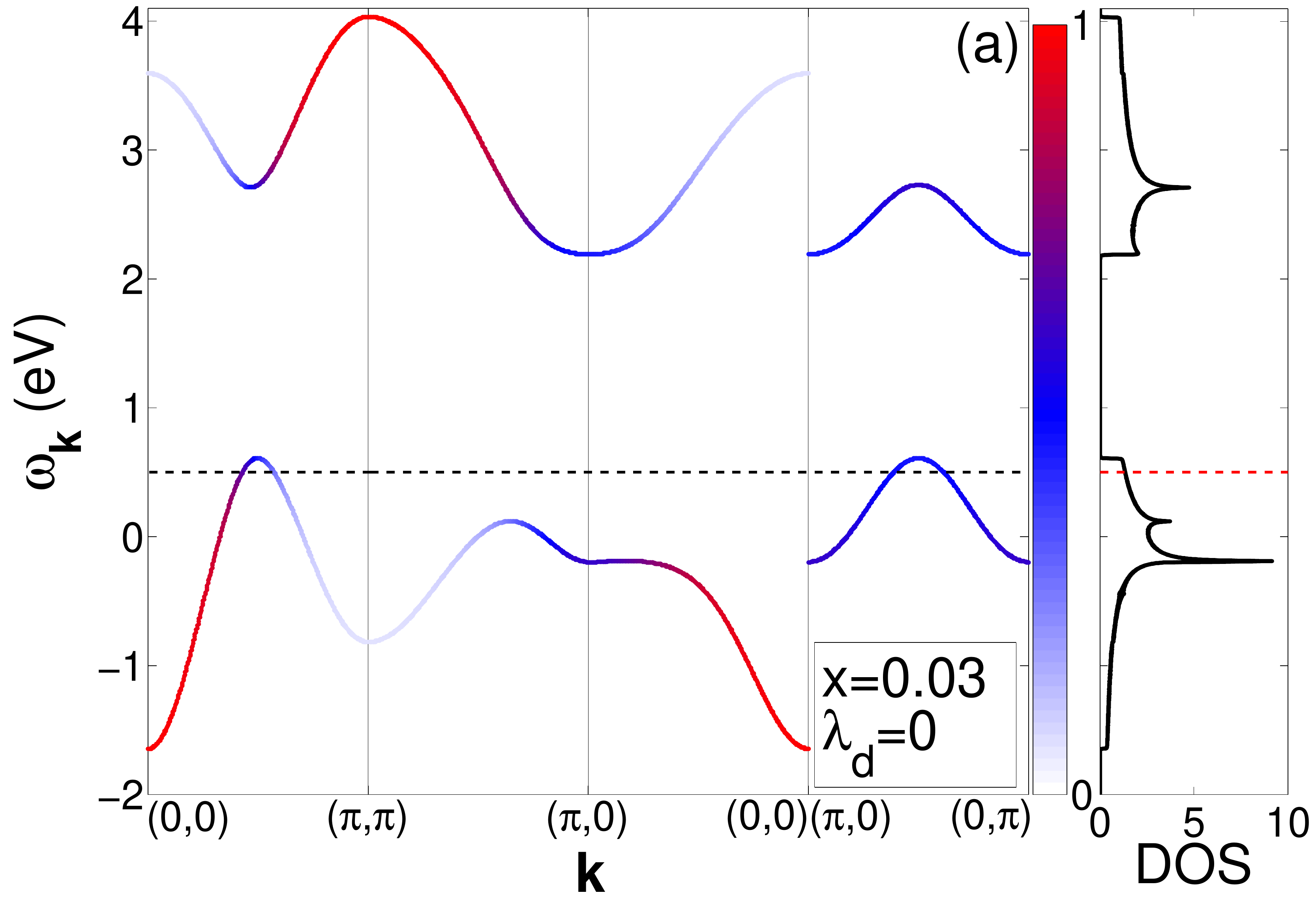}
\includegraphics[width=0.45\linewidth]{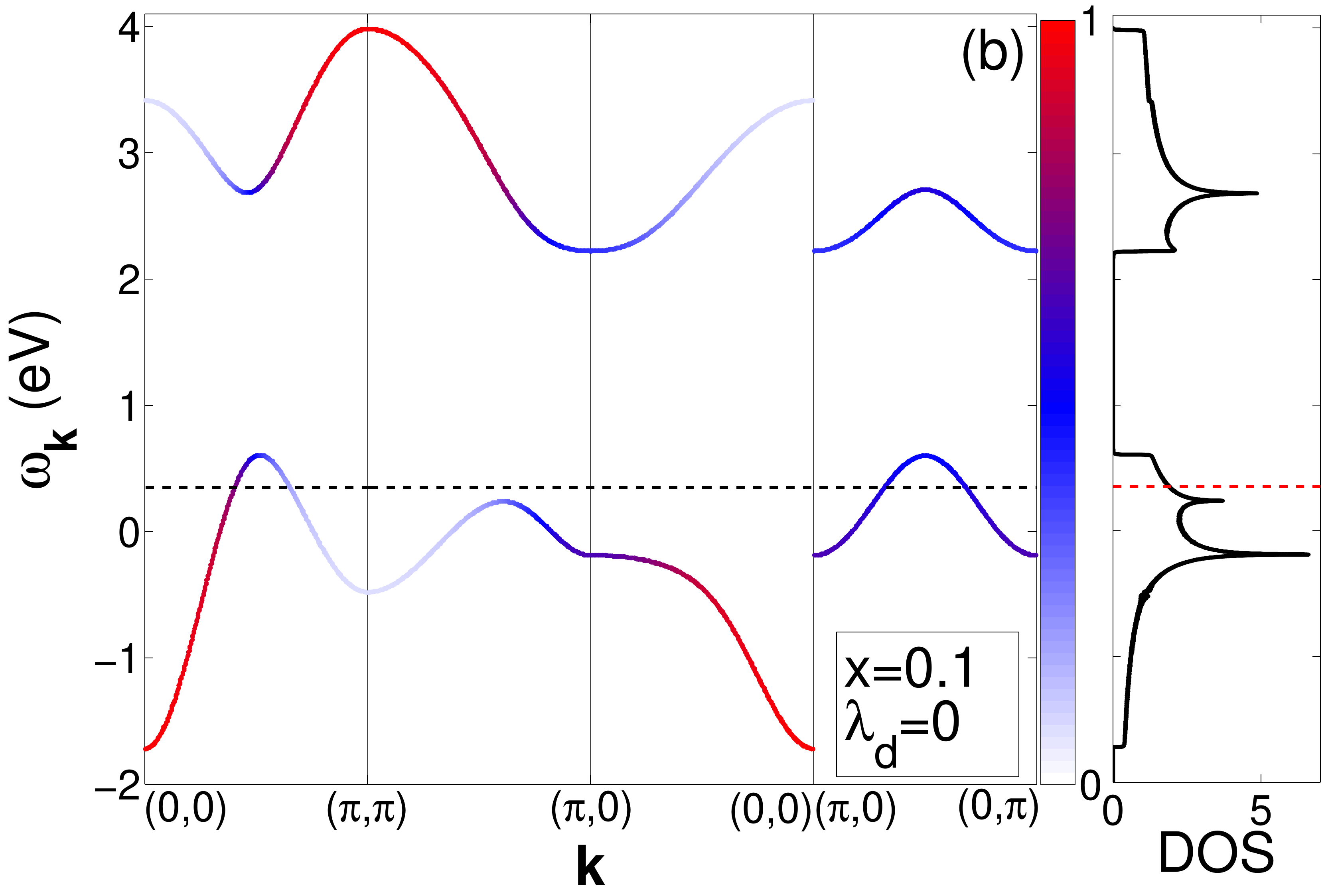}
\includegraphics[width=0.45\linewidth]{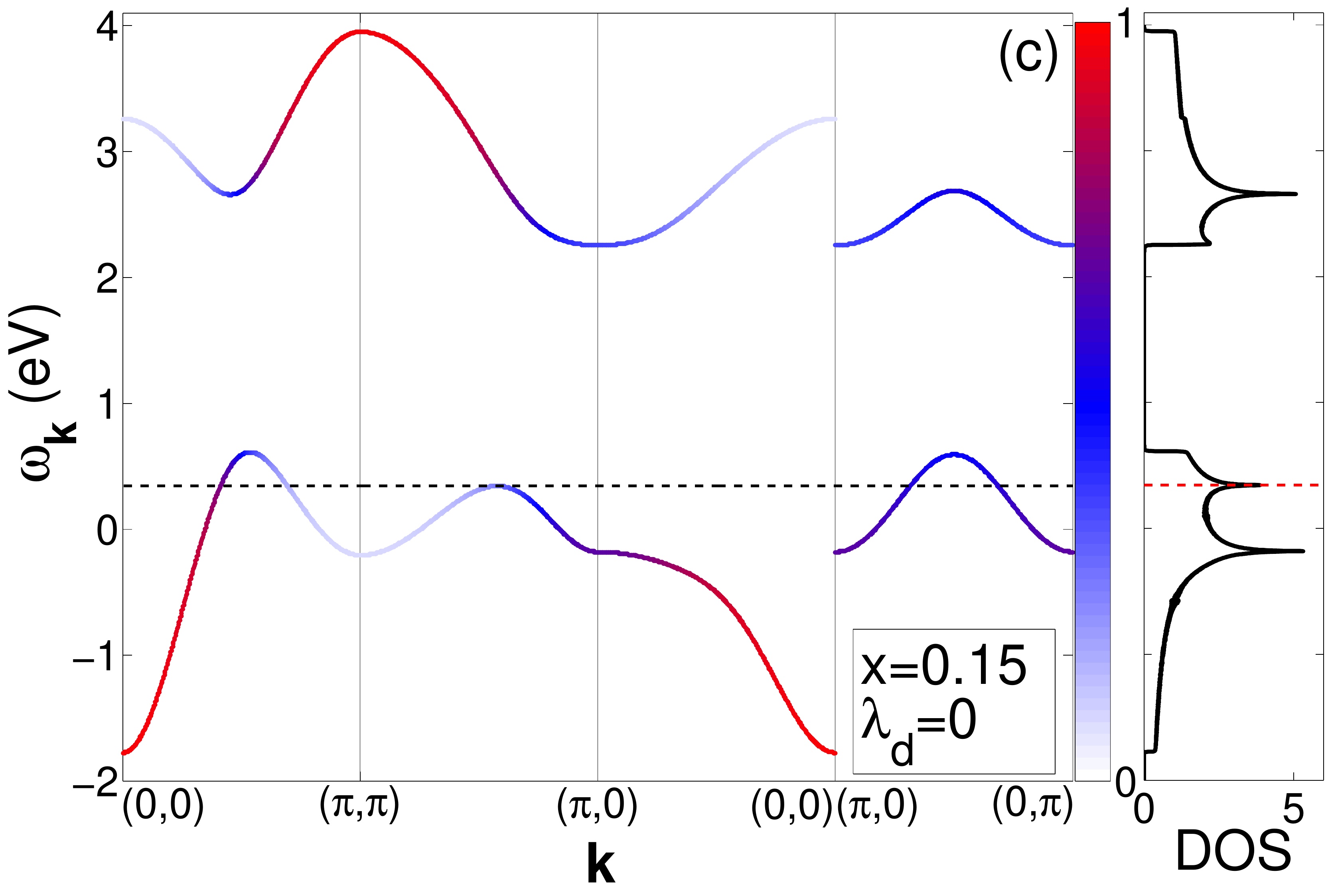}
\includegraphics[width=0.45\linewidth]{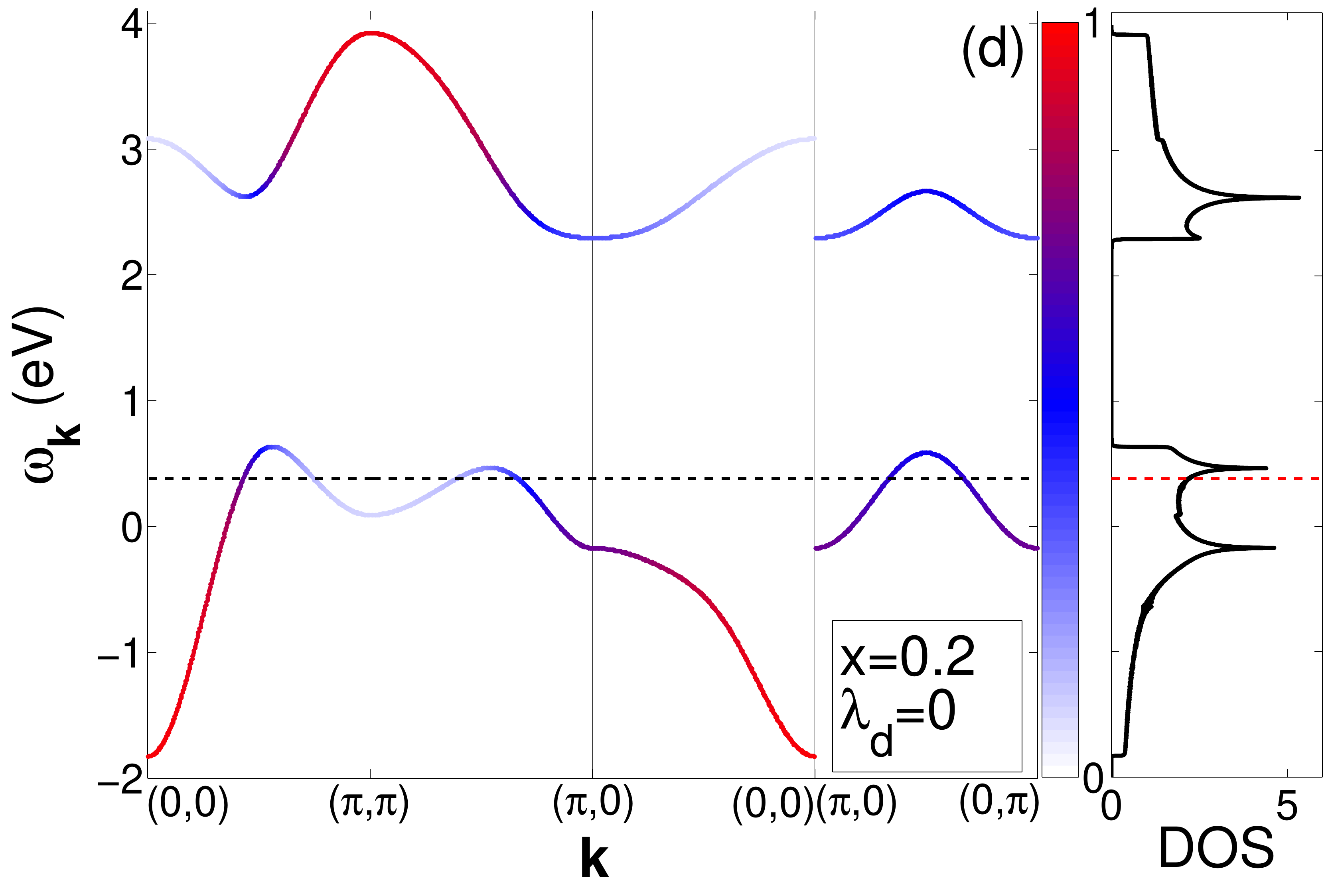}
\includegraphics[width=0.45\linewidth]{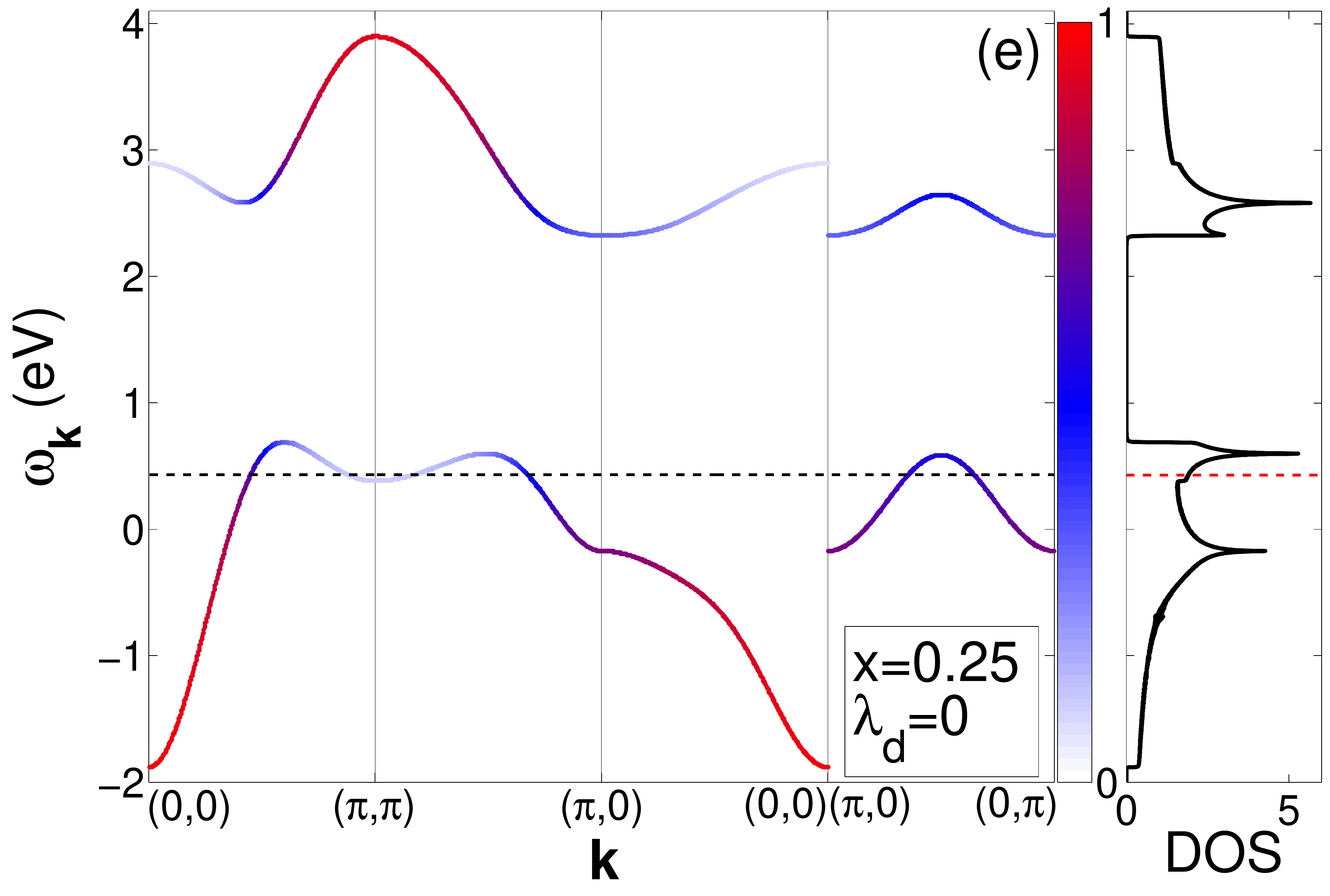}
\includegraphics[width=0.45\linewidth]{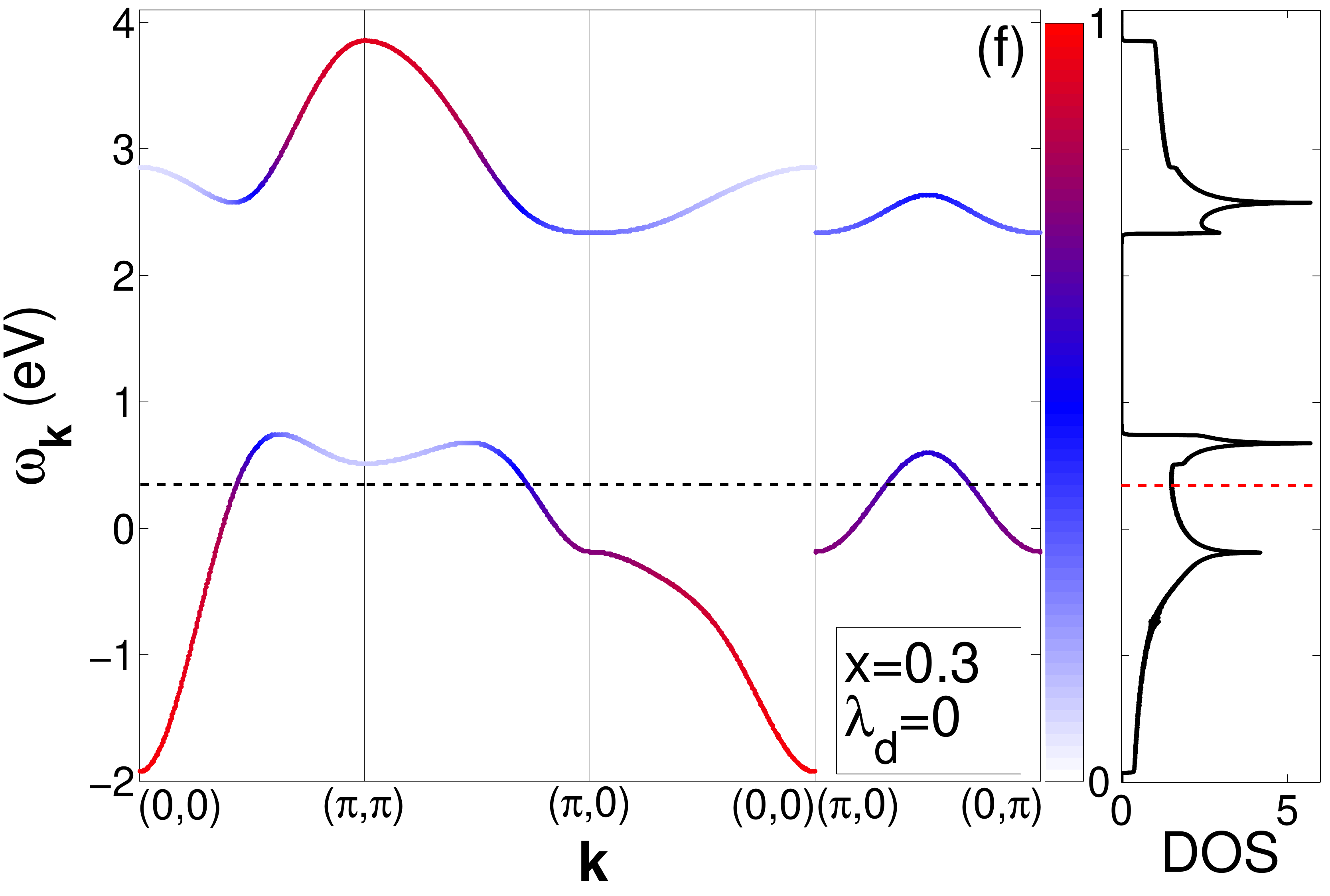}
\caption{\label{fig:bandstr_xdep_withoutEPI} Evolution of the band structure and density of states (DOS) of the quasiparticle excitations with doping $x$ for the pure electronic system (without phonon subsystem and EPI) within the frameworks of three-band $p-d$ model. (a) $x = 0.03$, (b) $x=0.1$, (c) $x=0.15$, (d) $x=0.2$, (e) $x=0.25$, (f) $x=0.3$). The conductivity (valence) bands are shown on the upper (lower) panel of each figure. Color in each $k$-point indicates the spectral weight of quasiparticle. Dashed line (black and red) shows Fermi level.}
\end{figure*}

\begin{figure}
\center
\includegraphics[width=1.0\linewidth]{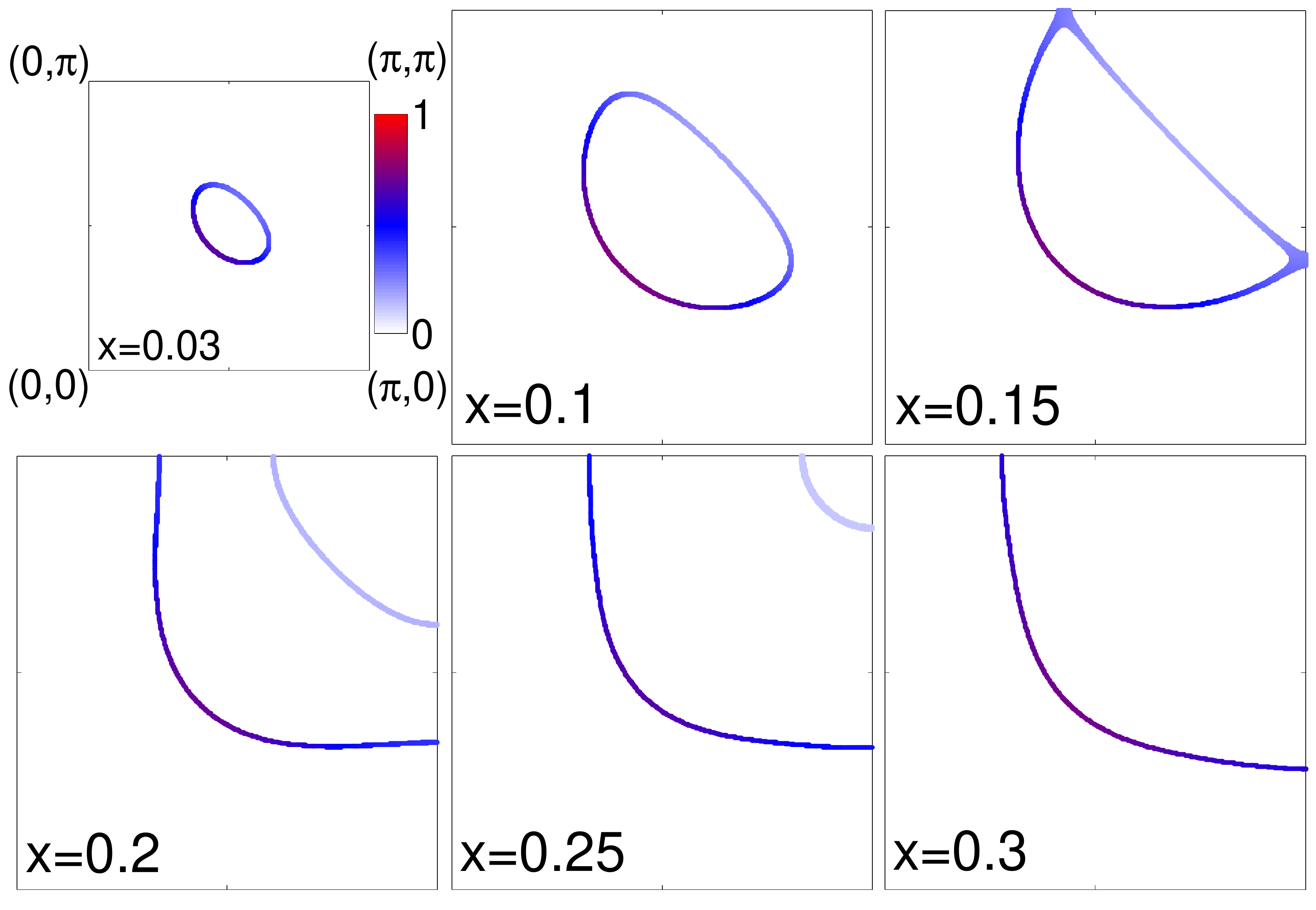}
\caption{\label{fig:fermisurf_xdep_withoutEPI} Evolution of the Fermi contour of the quasiparticle excitations with doping ($x = 0.03, 0.1, 0.15, 0.2, 0.25, 0.3$) for the pure electronic system (without phonon subsystem and EPI) within the frameworks of three-band $p-d$ model. Color in each k-point indicates the spectral weight of quasiparticle. Spectral weight is inhomogeneous in k-points along Fermi contour.}
\end{figure}

Filling of cluster eigenstates is determined self-consistently from the condition of completeness $\sum\limits_p {X_{\bf{f}}^{pp} = 1} $ and the chemical potential equation (here $n = 1 + x$ is the hole concentration for La$_{2-x}$Sr$_x$CuO$_4$):
\begin{eqnarray}
n &=& 1 + x = \nonumber \\
&=& \sum\limits_l {0 \cdot \left\langle {{X^{0l,0l}}} \right\rangle }  + \sum\limits_i {1 \cdot \left\langle {{X^{1i,1i}}} \right\rangle }  + \sum\limits_j {2 \cdot \left\langle {{X^{2j,2j}}} \right\rangle } \nonumber \\
\label{chempot_eq}
\end{eqnarray}
where $l,i,j$ runs over all zero-, single- and two-hole eigenstates respectively. Filling numbers for the zero-, single and two-hole states $\left\langle {{X^{0l,0l}}} \right\rangle$, $\left\langle {{X^{1i,1i}}} \right\rangle$, $\left\langle {{X^{2j,2j}}} \right\rangle$ are defined by the formulas
\begin{widetext}
\begin{equation}
\label{zap_nh0}
\left\langle {{X^{0l,0l}}} \right\rangle  = \left( { - \frac{1}{\pi }} \right)\frac{1}{N}\sum\limits_k {\int_{ - \infty }^\infty  {\frac{1}{{\exp \left( {{{\left( {\omega  - \mu } \right)} \mathord{\left/
 {\vphantom {{\left( {\omega  - \mu } \right)} {kT}}} \right.
 \kern-\nulldelimiterspace} {kT}}} \right) + 1}}{\mathop{\rm Im}\nolimits} {{\left\langle {\left\langle {{X_k^{0l,1i}}}
 \mathrel{\left | {\vphantom {{X_k^{0l,1i}} {X_k^{1i,0l}}}}
 \right. \kern-\nulldelimiterspace}
 {{X_k^{1i,0l}}} \right\rangle } \right\rangle }_{\omega  + i\delta }}} d\omega }
\end{equation}
\end{widetext}
\begin{widetext}
\begin{eqnarray}
\left\langle {{X^{1i,1i}}} \right\rangle  = \left( { - \frac{1}{\pi }} \right)\frac{1}{N}\sum\limits_k {\int_{ - \infty }^\infty  {\frac{1}{{\exp \left( {{{ - \left( {\omega  - \mu } \right)} \mathord{\left/
 {\vphantom {{ - \left( {\omega  - \mu } \right)} {kT}}} \right.
 \kern-\nulldelimiterspace} {kT}}} \right) + 1}}{\mathop{\rm Im}\nolimits} {{\left\langle {\left\langle {{X_k^{0l,1i}}}
 \mathrel{\left | {\vphantom {{X_k^{0l,1i}} {X_k^{1i,0l}}}}
 \right. \kern-\nulldelimiterspace}
 {{X_k^{1i,0l}}} \right\rangle } \right\rangle }_{\omega  + i\delta }}} d\omega }
\end{eqnarray}
\end{widetext}
\begin{widetext}
\begin{eqnarray}
\left\langle {{X^{2j,2j}}} \right\rangle  = \left( { - \frac{1}{\pi }} \right)\frac{1}{N}\sum\limits_k {\int_{ - \infty }^\infty  {\frac{1}{{\exp \left( { - {{\left( {\omega  - \mu } \right)} \mathord{\left/
 {\vphantom {{\left( {\omega  - \mu } \right)} {kT}}} \right.
 \kern-\nulldelimiterspace} {kT}}} \right) + 1}}{\mathop{\rm Im}\nolimits} {{\left\langle {\left\langle {{X_k^{1i,2j}}}
 \mathrel{\left | {\vphantom {{X_k^{1i,2j}} {X_k^{2j,1i}}}}
 \right. \kern-\nulldelimiterspace}
 {{X_k^{2j,1i}}} \right\rangle } \right\rangle }_{\omega  + i\delta }}} d\omega }
\end{eqnarray}
\end{widetext}

Note that filling numbers in zero-hole and two-hole sectors $\left\langle {{X^{0l,0l}}} \right\rangle$ and $\left\langle {{X^{2j,2j}}} \right\rangle$ are nonzero even in the undoped compound ($x = 0$), they are of the order of ${{{{\tilde t}^2}} \mathord{\left/
 {\vphantom {{{{\tilde t}^2}} {U_d^2}}} \right.
 \kern-\nulldelimiterspace} {U_d^2}}$. This fact is caused by hybridization of quasiparticle excitations between zero- and single-hole states and excitations between single- and two-hole states. In the presence of interband hoppings each of the upper and lower Hubbard bands are formed by superposition of excitations between states of zero- and single-hole Hilbert space sectors and between states of single- and two-hole Hilbert space sectors.

Here we will investigate electronic structure in interval of doping from $x = 0.03$ to $0.3$ . The short-range antiferromagnetic order is strong in the quasi-two-dimensional magnetic subsystem of La$_2$CuO$_4$ for wide region of doped hole concentration. We will describe the short-range order in the paramagnetic state as the isotropic spin liquid with zero spin projections and non-zero spin-spin correlations functions following to Refs.~\onlinecite{Shimahara1991,Barabanov1994,Valkov2005}. Spin-spin correlation functions for different doping levels was taken from the paper Ref.~\onlinecite{KorshOvch2007}.

\section{Doping dependence of the electronic structure within the three-band p-d model\label{x_dep_without_EPI}}
\begin{figure*}
\center
\includegraphics[width=0.45\linewidth]{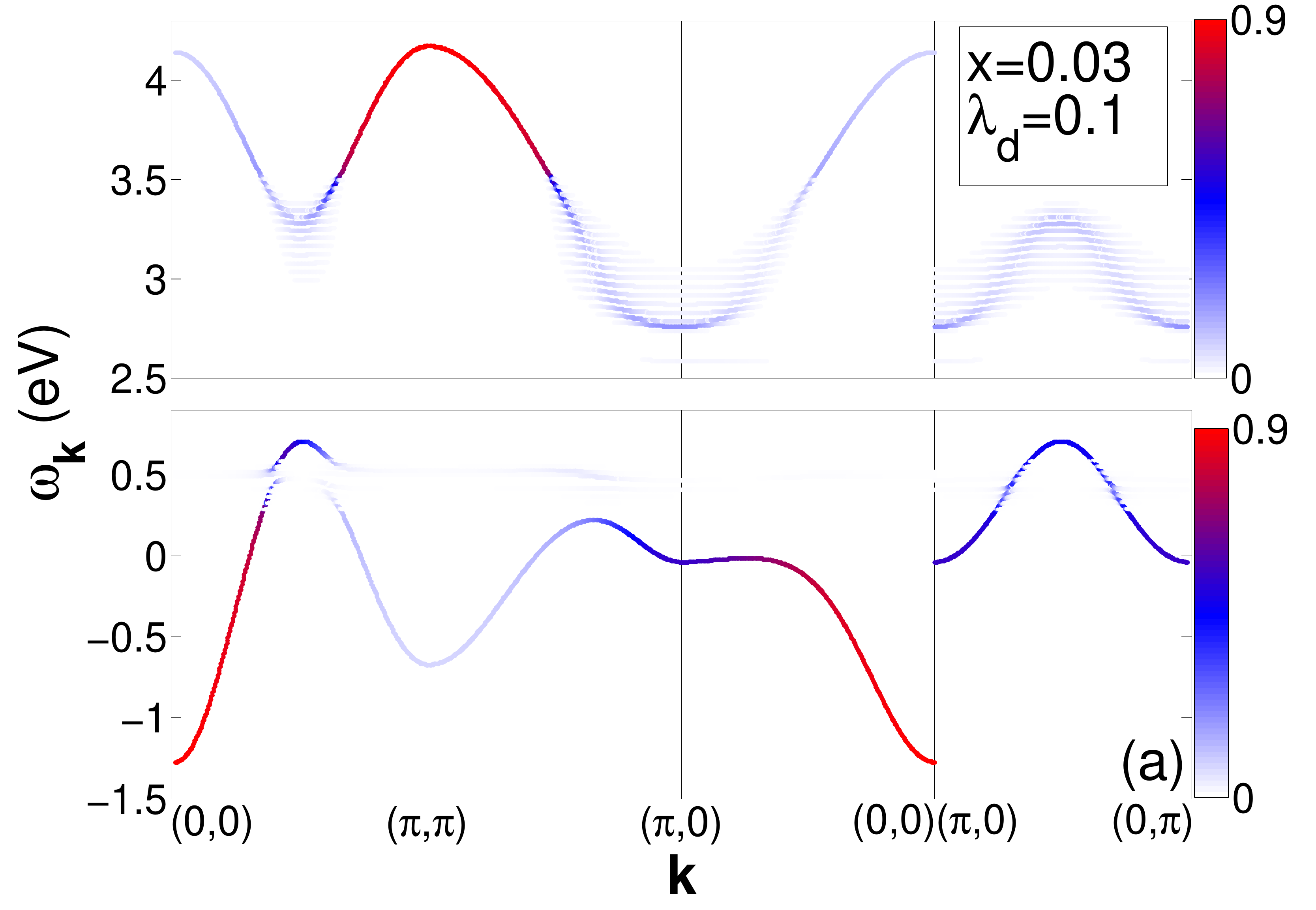}
\includegraphics[width=0.45\linewidth]{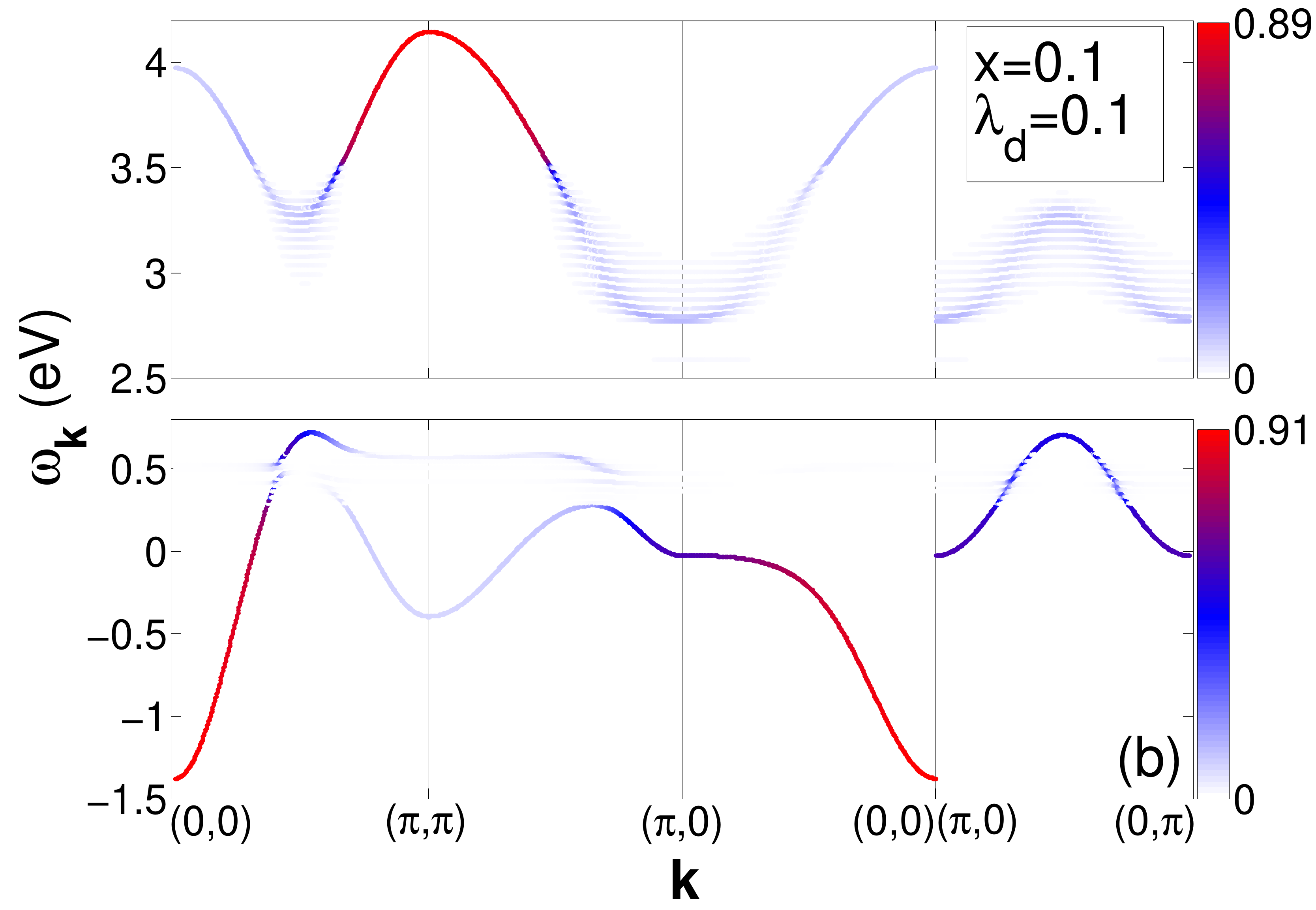}
\includegraphics[width=0.45\linewidth]{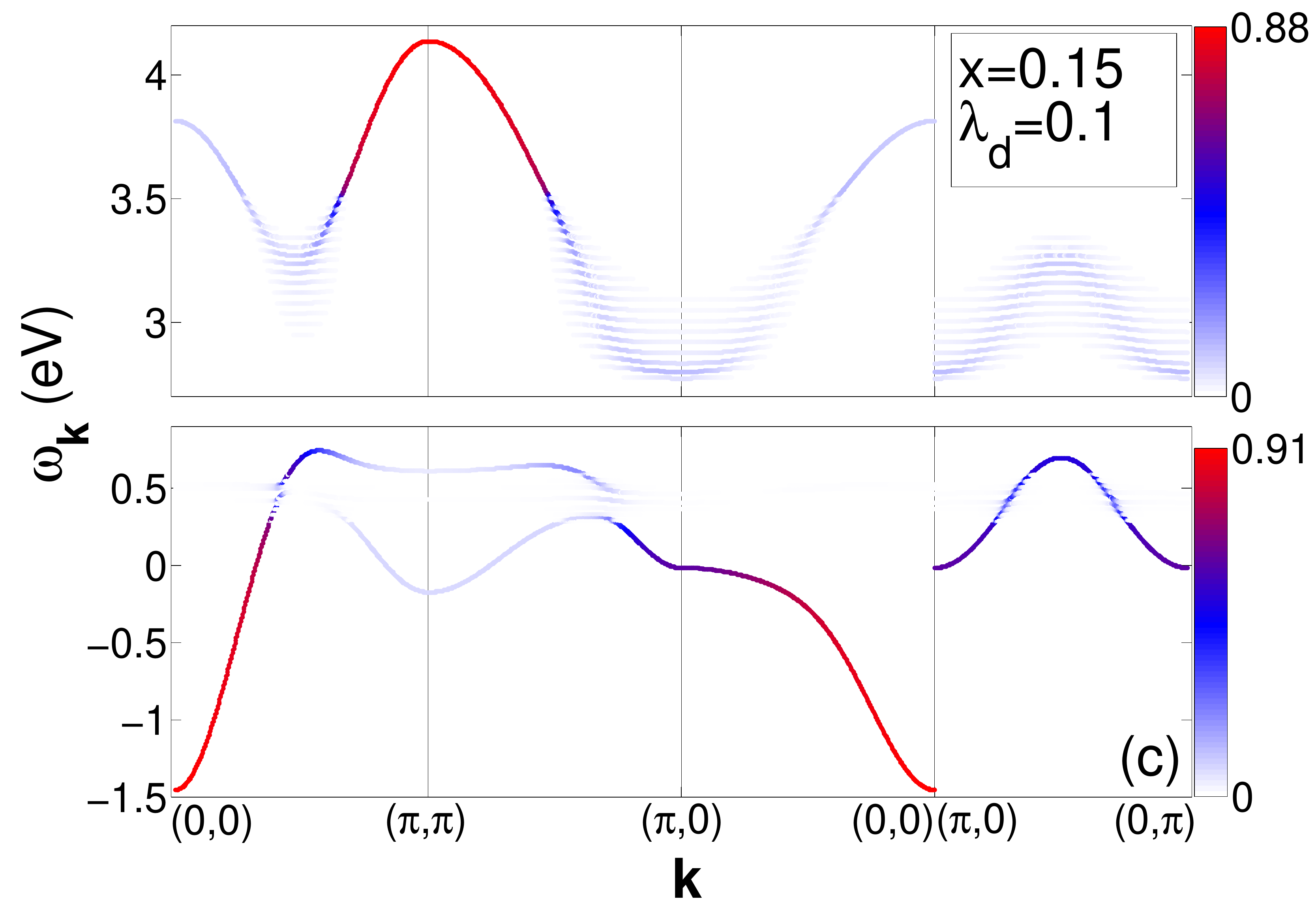}
\includegraphics[width=0.45\linewidth]{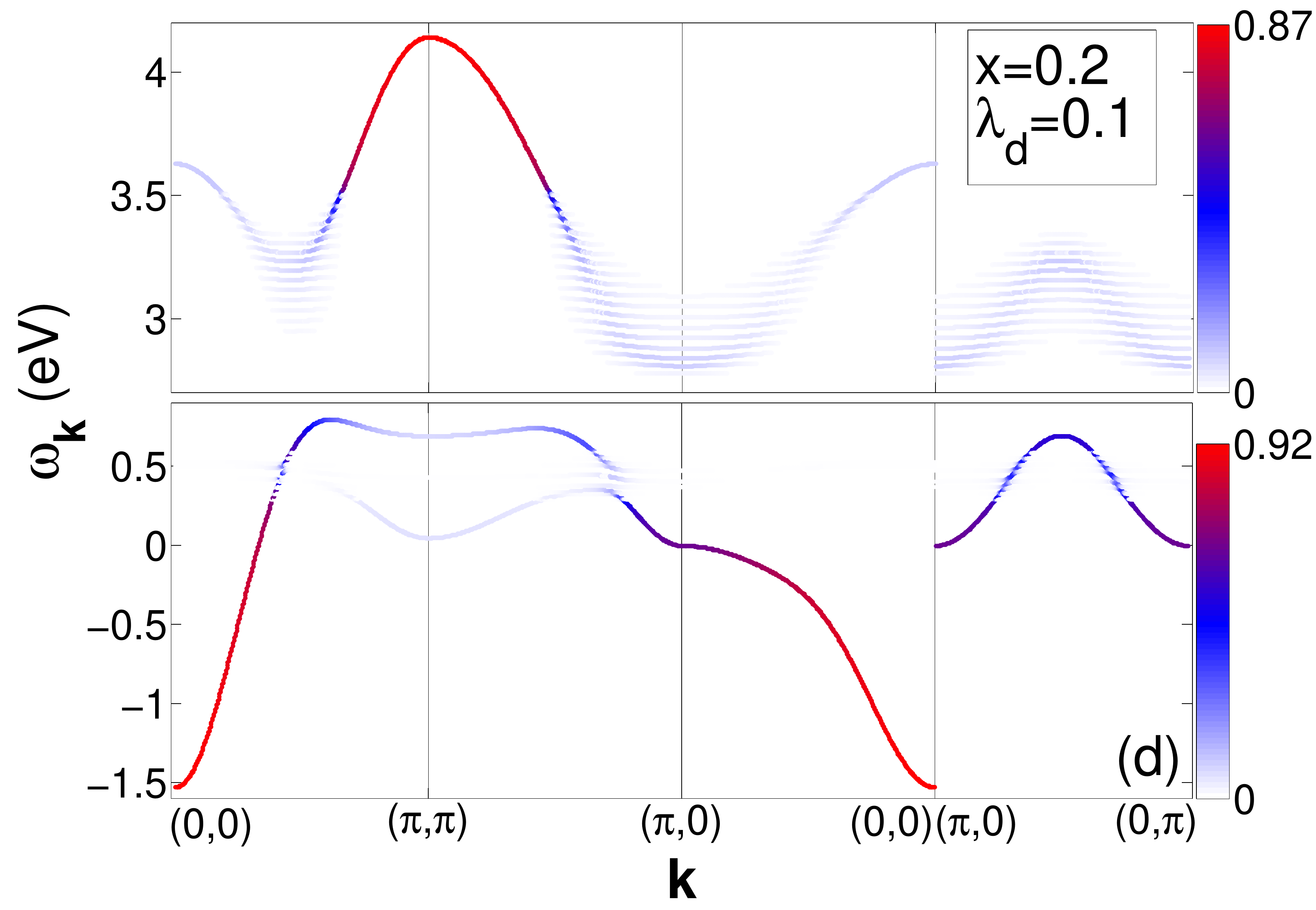}
\includegraphics[width=0.45\linewidth]{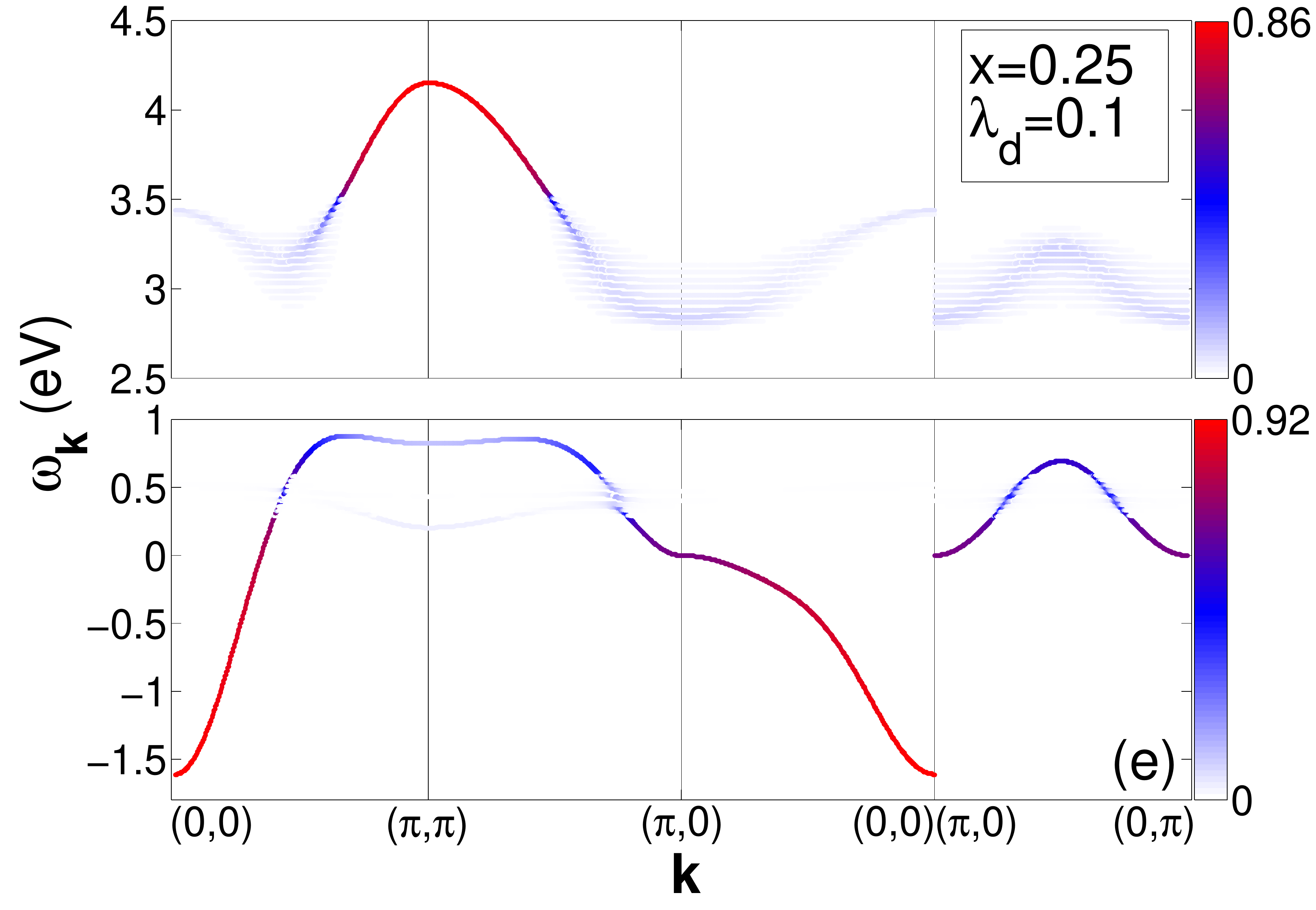}
\includegraphics[width=0.45\linewidth]{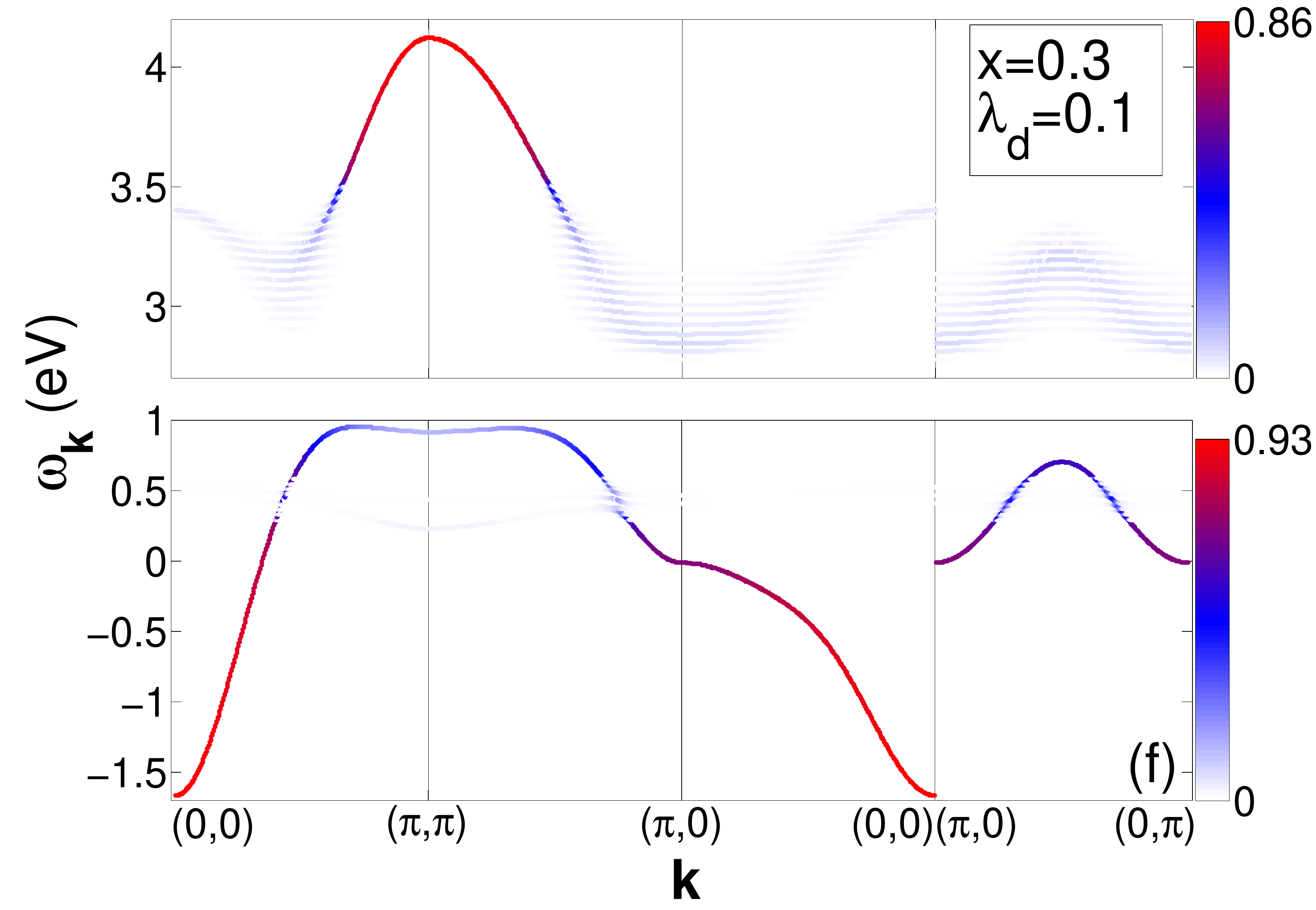}
\caption{\label{fig:elstr_xdep_withEPI} Evolution of band structure with doping at EPI constant ${\lambda _d} = 0.1$ and temperature $T=10$ K. (a) $x=0.03$, (b) $x=0.1$ (underdoped); (c) $x=0.15$ (optimally doped); (d) $x=0.2$, (e) $x=0.25$, (f) $x=0.3$ (overdoped). Upper part of each subfigure shows conductivity band, lower part - valence band. Color at each point of band structure displays spectral weight at this point of $k$-space. Color scale of spectral weight differs for valence and conductivity bands. Unit on common scale corresponds to maximal intensity of quasiparticle in the system with phonon subsystem and EPI.}
\end{figure*}
\begin{figure*}
\center
\includegraphics[width=0.45\linewidth]{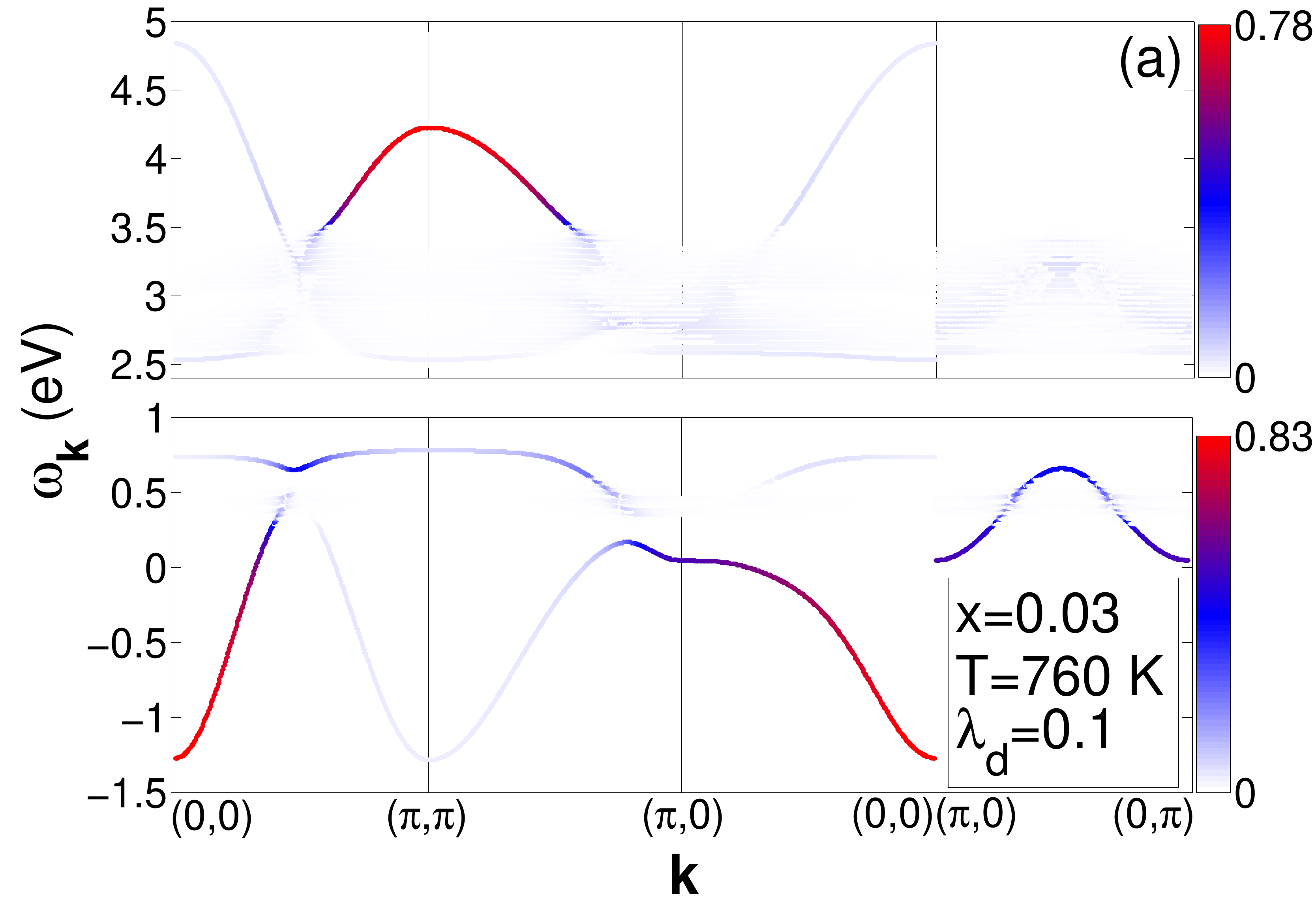}
\includegraphics[width=0.45\linewidth]{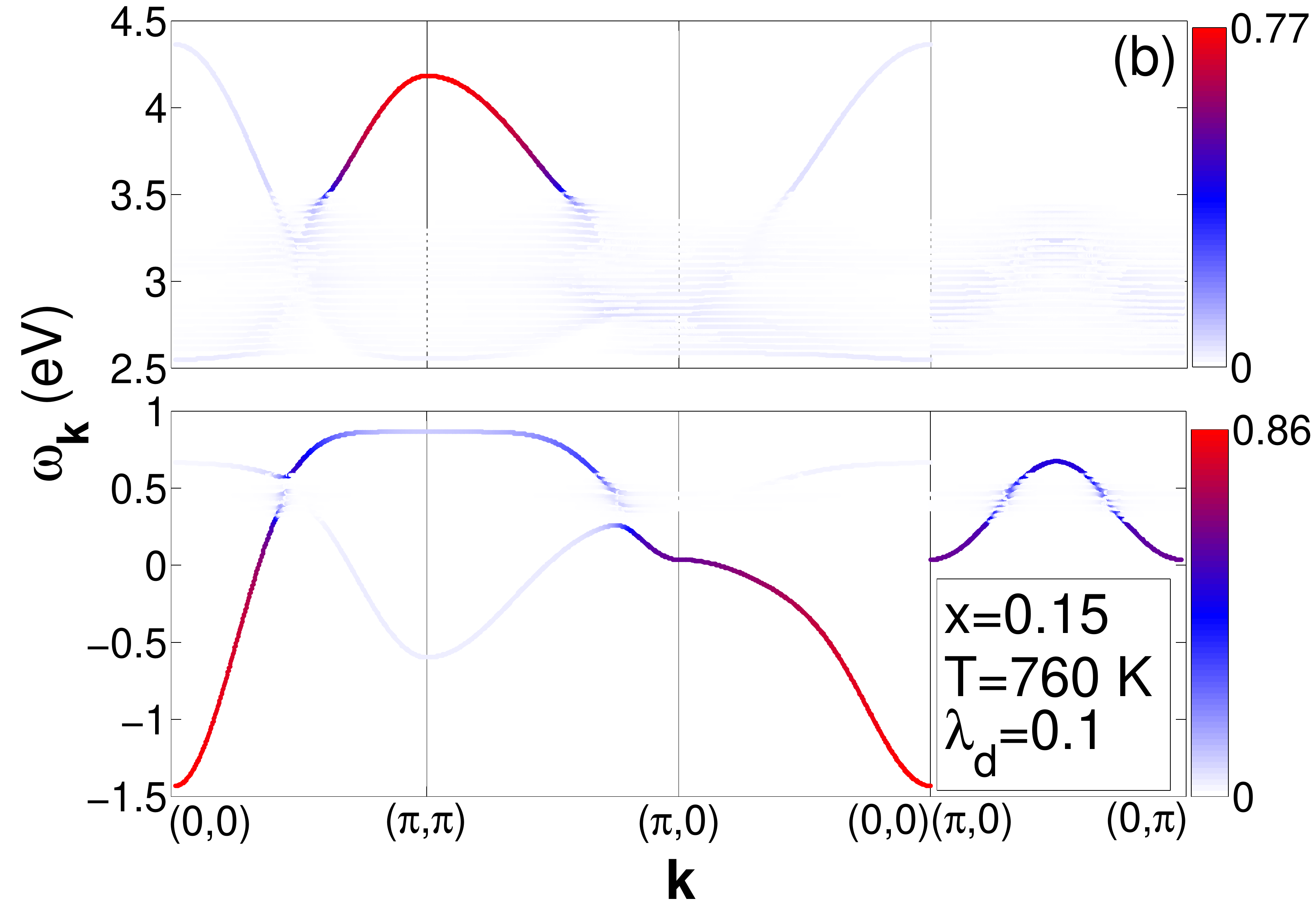}
\includegraphics[width=0.45\linewidth]{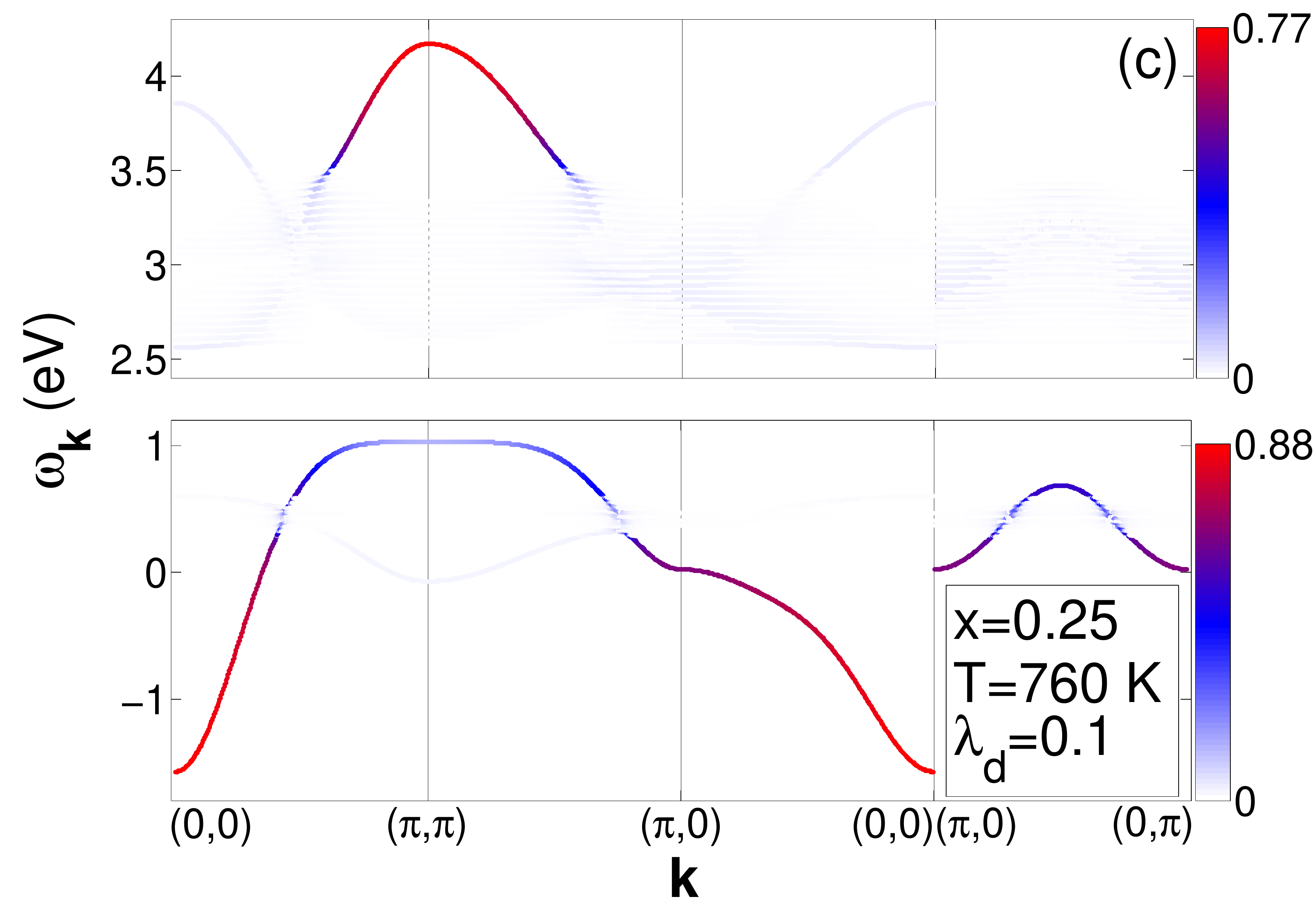}
\caption{\label{fig:elstr_x_T_dep_withEPI} Evolution of band structure with doping at EPI constant ${\lambda _d} = 0.1$ and high temperature $T=760$ K. (a) $x=0.03$ (underdoped); (b) $x=0.15$ (optimally doped); (c) $x=0.25$ (overdoped). Upper part of each subfigure shows conductivity band, lower part - valence band. Color at each point of band structure displays spectral weight at this point of $k$-space. Color scale of spectral weight differs for valence and conductivity bands. Unit on common scale corresponds to maximal intensity of quasiparticle in the system with phonon subsystem and EPI.}
\end{figure*}
Without EPI region of the low-energy excitations of upper and lower Hubbard bands is formed by two quasiparticles. Each quasiparticle is the superposition of excitations between ground zero- and single-hole states and between ground single- and two-hole states. It is seen from Fig.~\ref{fig:bandstr_xdep_withoutEPI}(a)-(f) that spectral weight (it is shown by color at each point of k-space) is inhomogeneously distributed over dispersion surface. For LHB spectral weight is maximal at point ${\bf{k}} = \left( {0,0} \right)$, decreases with increasing ${\bf{k}}$ up to the boundaries of Brillouin zone. Inverse tendency for the UHB, spectral weight is maximal at point ${\bf{k}} = \left( {\pi ,\pi } \right)$ and minimal is at ${\bf{k}} = \left( {0,0} \right)$. Inhomogeneous distribution of spectral weight is caused by interband hopping integral $\tilde t$ between Hubbard bands formed due to strong electronic correlations. This is the reason of inhomogeneous distribution of spectral weight over Fermi contour for all doping levels (Fig.~\ref{fig:fermisurf_xdep_withoutEPI}). Maximal spectral weight is in the nodal direction on the side of hole pocket with ${\bf{k}}$ smaller than ${\bf{k}} = \left( {{\pi  \mathord{\left/
 {\vphantom {\pi  2}} \right.
 \kern-\nulldelimiterspace} 2},{\pi  \mathord{\left/
 {\vphantom {\pi  2}} \right.
 \kern-\nulldelimiterspace} 2}} \right)$ at the opposite side of pocket (with ${\bf{k}}$ larger than ${\bf{k}} = \left( {{\pi  \mathord{\left/
 {\vphantom {\pi  2}} \right.
 \kern-\nulldelimiterspace} 2},{\pi  \mathord{\left/
 {\vphantom {\pi  2}} \right.
 \kern-\nulldelimiterspace} 2}} \right)$) spectral weight is depressed. Reduced intensity of quasiparticle excitations at Fermi contour may be reason of its absence in the ARPES Fermi surfaces. Arc can be high intensity part of hole pocket.
 At small concentration of doped holes Fermi contour is small hole pocket centered around ${\bf{k}} = \left( {{\pi  \mathord{\left/
 {\vphantom {\pi  2}} \right.
 \kern-\nulldelimiterspace} 2},{\pi  \mathord{\left/
 {\vphantom {\pi  2}} \right.
 \kern-\nulldelimiterspace} 2}} \right)$ (Fig.~\ref{fig:fermisurf_xdep_withoutEPI},$x=0.03$). Growth of hole concentration leads to reconstruction of band structure and Fermi contour wherein Fermi level shifts deeper to LHB. LHB band is modified more than UHB. The main effect of doping is defined by the increase of energy of LHB quasiparticles at point ${\bf{k}} = \left( {\pi ,\pi } \right)$. Hole pocket becomes larger with doping. At ${\bf{k}} = \left( {\pi ,\pi } \right)$ Fermi level touches another part of dispersion surface at point of direction ${\bf{k}} = \left( {\pi ,0} \right)$ - ${\bf{k}} = \left( {\pi ,\pi } \right)$ (${\bf{k}} = \left( {0,\pi } \right)$ - ${\bf{k}} = \left( {\pi ,\pi } \right)$) (Fig.~\ref{fig:fermisurf_xdep_withoutEPI},$x=0.15$). It results in quantum phase transition which is accompanied by closing of four hole pockets in the full Brillouin zone with its transformation into two large contours, bigger one is hole contour, smaller one have electron type (Fig.~\ref{fig:fermisurf_xdep_withoutEPI},$x=0.2$). Spectral weight is different in the hole and electron contours. Further doping leads to growth of the hole contour and the reduction of electron contour (Fig.~\ref{fig:fermisurf_xdep_withoutEPI},$x=0.25$). Nearly at $x = 0.26$ when Fermi level crosses local minimum at point ${\bf{k}} = \left( {\pi ,\pi } \right)$ second quantum phase transition occurs: electron contour disappears  and only hole pocket remains (Fig.~\ref{fig:fermisurf_xdep_withoutEPI},$x=0.3$).

\section{Doping dependence of the electronic structure of the polaron quasiparticles \label{x_dep_with_EPI}}

Occupation of the ground two-hole local polaron state grows and quasiparticle excitations formed by this state acquire spectral weight with increasing concentration of doped holes. Consequently new polaron quasiparticle bands appear in the system with phonon subsystem and EPI (Fig.~\ref{fig:levels}). Generally polaron band structure results from hybridization of Hubbard fermions (quasiparticle excitation $0-0$, transition between ground single- and two-hole states) with Franck-Condon resonances. Dispersion constructed over maxima of spectral function even at large EPI qualitatively coincides with dispersion of pure electronic system. Splittings due to hybridization with Franck-Condon resonances in the system with doped holes appear (Fig.~\ref{fig:elstr_xdep_withEPI}(a)-(f)) in the modified valence band of Hubbard fermions. Two tendencies of band structure reconstruction can be distinguished with doping. Firstly reconstruction of Hubbard fermions band is in the form of the top of the valence band splitting off. With further doping reconstruction of the top of the valence band as its low-energy part is similar to reconstruction of LHB without EPI: energy of the local minimum at point ${\bf{k}} = \left( {\pi ,\pi } \right)$ increases (Fig.~\ref{fig:elstr_xdep_withEPI}(a)-(f)). Secondly number of splittings due to hybridization of the new bands with wide band of $0-0$ excitation increases, splittings are at energies which are larger than degenerate phononless resonances $0-0$, $1-1$, $2-2$ etc (Fig.~\ref{fig:FC_resonances}). Sequence of the Fermi contour transformations with doping is qualitatively preserved for moderate EPI constants since LHB is slightly affected by the polaron subbands splittings.

\section{Temperature dependence of the electronic structure of the polaron quasiparticles in the doped compounds \label{T_x_dep_with_EPI}}

It is seen from temperature and concentration dependences of the band structure of undoped cuprates that top of the valence band splitting off appears as with temperature increasing as with doping (Fig.~\ref{fig:bandstr_xdep_withoutEPI},Fig.~\ref{fig:elstr_xdep_withEPI}). It results from hybridization of $0-0$ quasiparticles with new quasiparticle excitations which acquire spectral weight and dispersion with filling excited local single-hole states or ground two-hole state. Temperature influences on top of the valence band and almost doesn't change low-energy part of LHB in the system with doped holes. Temperature growth in the underdoped system modifies top of the valence band by the same way as in the system without doped holes: local maximum at point ${\bf{k}} = \left( {{\pi  \mathord{\left/
 {\vphantom {\pi  2}} \right.
 \kern-\nulldelimiterspace} 2},{\pi  \mathord{\left/
 {\vphantom {\pi  2}} \right.
 \kern-\nulldelimiterspace} 2}} \right)$ is everted and becomes local minimum, maximum energy transfers to point ${\bf{k}} = \left( {\pi ,\pi } \right)$ (Fig.~\ref{fig:elstr_x_T_dep_withEPI}(a)). Spectral weight is homogeneously distributed over all $k$-points, dispersion becomes weaker as compared with $T=0$ case, flat bands are formed around points ${\bf{k}} = \left( {0 ,0 } \right)$ and ${\bf{k}} = \left( {\pi ,\pi } \right)$ (Fig.~\ref{fig:elstr_x_T_dep_withEPI}(a)). For optimal doping at high temperature ($T=760$ K) local minimum of splitted off top of the valence band at ${\bf{k}} = \left( {\pi ,\pi } \right)$ disappears, instead of this flat band is formed (Fig.~\ref{fig:elstr_x_T_dep_withEPI}(b)). For overdoped system at temperature $T=760$ K global maximum of the valence band is formed at ${\bf{k}} = \left( {\pi ,\pi } \right)$ and dispersion of the spectral function maximum is similar to dispersion of LHB in the paramagnetic phase within three-band $p-d$ model (Fig.~\ref{fig:elstr_x_T_dep_withEPI}(c)).

\section{Conclusion \label{Summary}}
We obtain doping dependence of the band structure, Fermi contour and density of states of quasiparticle excitations within the three-band $p-d$ model. Evolution of Fermi contour from small hole pockets to large hole contour is accompanied by two quantum phase transition. Taking into account interband hoppings in considered model results in inhomogeneous spectral weight along Fermi contour. Small spectral weight in the one side of pocket may be invisible in ARPES experiments. Thus it is possible that there is no contradiction between "arc" and "pockets" viewpoints.

\begin{acknowledgments}
Authors are thankful to Russian Science Foundation (project No. 14-12-00061) for financial support.
\end{acknowledgments}

\bibliography{C:/literature_x_T_dep_pol}

\end{document}